\documentclass[12pt,preprint]{aastex}
\usepackage{underscore}
\usepackage{color}
\usepackage{lscape}

\slugcomment{Accepted for publication in the Astronomical Journal}

\shorttitle{15P/Finlay} 
\shortauthors{Ishiguro et al.}

\begin{document}

\title{2014--2015 Multiple Outbursts of 15P/Finlay}

\author{Masateru ISHIGURO}
\affil{Department of Physics and Astronomy, Seoul National University,\\
Gwanak, Seoul 151--742, South Korea}
\email{ishiguro@astro.snu.ac.kr}

\author{Daisuke KURODA}
\affil{Okayama Astrophysical Observatory, National Astronomical
Observatory of Japan, Asakuchi, Okayama 719--0232, Japan}

\author{Hidekazu HANAYAMA}
\affil{Ishigakijima Astronomical Observatory, National Astronomical 
Observatory of Japan, 1024--1 Arakawa, Ishigaki, Okinawa 907--0024, Japan}

\author{Yuna Grace Kwon, Yoonyoung KIM, Myung Gyoon LEE}
\affil{Department of Physics and Astronomy, Seoul National University,
Gwanak, Seoul 151--742, South Korea}

\author{Makoto WATANABE}
\affil{Faculty of Science Department of Applied Physics, Okayama University of Science, 1-1 Ridai-cho, Okayama-si, Okayama 700-0005, Japan}

\author{Hiroshi AKITAYA, Koji KAWABATA, Ryosuke ITOH, Tatsuya NAKAOKA, Michitoshi YOSHIDA}
\affil{Hiroshima Astrophysical Science Center, Hiroshima University, Higashihiroshima, Hiroshima 739-8526, Japan }

\author{Masataka IMAI}
\affil{Department of Cosmosciences, Graduate School of Science, Hokkaido University, Kita-ku, Sapporo 060-0810, Japan}

\author{Yuki SARUGAKU}
\affil{Kiso Observatory, Institute of Astronomy, Graduate School of Science, The University of Tokyo,
Mitake, Kiso-machi, Kiso, Nagano, 397-0101, Japan}

\author{Kenshi YANAGISAWA}
\affil{Okayama Astrophysical Observatory, National Astronomical
Observatory of Japan, Asakuchi, Okayama 719-0232, Japan}

\author{Kouji OHTA}
\affil{Department of Astronomy, Kyoto University, Kyoto 606-8502, Japan}

\author{Nobuyuki KAWAI}
\affil{Department of Physics, Tokyo Institute of Technology
2-12-1 Ookayama, Meguro-ku, Tokyo 152-8551, Japan}

\author{Takeshi MIYAJI}
\affil{Ishigakijima Astronomical Observatory, National Astronomical
Observatory of Japan,\\ Ishigaki, Okinawa 907-0024, Japan}

\author{Hideo FUKUSHIMA}
\affil{National Astronomical Observatory of Japan, Mitaka, Tokyo,
181-8588, Japan} 

\author{Satoshi HONDA, Jun TAKAHASHI}
\affil{Nishi-Harima Astronomical Observatory, Center for Astronomy,
University of Hyogo, Sayo, Hyogo 679-5313, Japan}

\author{Mikiya SATO}
\affil{Kawasaki Municipal Science Museum, Kawasaki, Kanagawa 214-0032, Japan} 

\author{Jeremie J. VAUBAILLON}
\affil{Observatoire de Paris, I.M.C.C.E., Denfert Rochereau, Bat. A.,
F-75014 Paris, France}

\author{Jun-ichi WATANABE}
\affil{National Astronomical Observatory of Japan, Mitaka, Tokyo,
181-8588, Japan}

\begin{abstract}
Multiple outbursts of a Jupiter-family comet, 15P/Finlay,  occurred from late 2014 to early 2015.
We conducted an observation of the comet after the first outburst and subsequently
witnessed another outburst on 2015 January 15.6--15.7. The gas, consisting mostly of C$_{2}$ and CN, and
dust particles expanded at speeds of 1,110 $\pm$ 180 m s$^{-1}$ and 570 $\pm$ 40 m s$^{-1}$ at a heliocentric distance of 1.0 AU.
We estimated the maximum ratio of solar radiation pressure with respect to the solar gravity  $\beta_\mathrm{max}$ = 1.6 $\pm$ 0.2,
which is consistent with porous dust particles composed of silicates and organics. We found that 
10$^8$--10$^9$ kg of dust particles (assumed to be 0.3 \micron--1 mm) were ejected through each outburst.
Although the total mass is three orders of magnitude smaller
than that of the 17P/Holmes event observed in 2007, the kinetic energy per unit mass (10$^4$ J kg$^{-1}$) is equivalent to the estimated values of
17P/Holmes and 332P/2010 V1 (Ikeya--Murakami), suggesting that the outbursts were caused by a similar physical mechanism.
From a survey of cometary outbursts on the basis of voluntary reports, we conjecture that 15P/Finlay--class outbursts
occur $>$1.5 times annually and inject dust particles from Jupiter-family comets and Encke-type comets into
interplanetary space at a rate of $\sim$10 kg s$^{-1}$ or more.
\end{abstract}



\section{INTRODUCTION}
\label{sec:introduction}

15P/Finlay (hereafter 15P) was an undistinguished comet discovered by William Henry Finlay at Cape of Good Hope,
South Africa, on 1886 September 26. This comet has a semi-major axis of $a$ = 3.488 AU, eccentricity of $e$ = 0.720, inclination of $i$ = 6.80\arcdeg, and
Tisserand parameter with respect to Jupiter of $T_\mathrm{J}$ = 2.62, which are typical of Jupiter-family comets (JFCs).
Since the discovery, it showed irregular magnitude light curves at different apparitions \citep{Sekanina1993}.
The effective radius of 15P is estimated to be 0.92 $\pm$ 0.05 km \citep{Fernandez2013}, which is consistent with
early results in \citet{Whipple1977} and \citet{Mendis1985}.
 It has maintained the perihelion around the Earth
orbit at 0.98--1.10 AU for about a century and is sometimes linked to a meteor shower \citep{Beech1999,Terentjeva2011}.
It is likely that the measured absolute magnitude reduced by a factor of $\sim$10 from 7.5 mag in 1886 to 10.1 mag in 1981 \citep{Kresak1989},
suggesting that 15P might have lost a  fraction of volatile components near the surface while developing a dust mantle layer on the surface
similar to other periodic comets \citep[e.g.,][]{Kwon2016,Hsieh2015}.

The comet exhibited two large-scale outbursts around the perihelion passage in 2014--2015,
with the first outburst occurring on 2014 December 16 \citep{Ye2015}.
The image showed an envelope feature and near--nuclear tail, which is reminiscent of past cometary outbursts at 17P/Holmes and 332P/2010 V1 (Ikeya--Murakami), hereafter referred to as 17P and 332P, respectively \citep{Ishiguro2010,Ishiguro2014}.
Soon after the report of the first outburst, we conducted an observation from 2014 December 23 to 2015 March 16
for deepening our understanding of cometary outbursts. We used six ground-based telescopes that
constitute a portion of the Optical and Infrared Synergetic Telescopes for Education and Research (OISTER) inter-university observation network.
As a result of frequent observation several times a week, we witnessed the second outburst on UT 2015 January 15.

Such cometary outbursts have drawn the attention of researchers because they offer insight into the physical properties of comet nuclei
\citep{Hughes1990}. The huge outburst of 17P could be explained by the
crystallization of buried amorphous ice \citep{Li2011}. Although similar morphological features were found at 332P \citep{Ishiguro2014},
numerous of fragments were identified at its return in 2016 \citep{Weryk2016,Kleyna2016}. Motivated by a series of detections
regarding cometary outbursts, we investigated the physical properties of the 15P multiple outbursts and estimated the frequency and
mass production rate of outbursts on a scale similar to 15P. We describe our observations and data analysis in Section 2,
and the photometric and polarimetric results in Section 3. We then discuss our findings considering the reports of recent outbursts
in Section 4.

\section{OBSERVATIONS AND DATA ANALYSIS}
\label{sec:observations}
The observational journal is summarized in Table 1.
Imaging observations were conducted from UT 2014 December 23 to UT 2015 March 16 using four telescopes: the
Okayama Astrophysical Observatory (OAO) 0.5-m reflecting robotic telescope (OAO 0.5 m) and 1.88-m telescope
(OAO 1.88 m), the Ishigakijima Astronomical Observatory (IAO) Murikabushi 1.05-m telescope (IAO 1.05 m), and the
Nishi--Harima Astronomical Observatory (NHAO) Nayuta 2-m telescope (NHAO 2 m).
We employed standard charge coupled device (CCD) cameras, that is, Multicolor Imaging Telescopes for Survey and Monstrous
Explosions (MITSuME) systems with Sloan Digital Sky Survey (SDSS) {\sl g}$'$,
Johnson--Cousins $R_\mathrm{C}$, and $I_\mathrm{C}$--band filters attached to OAO 0.5 m and IAO 1.05 m;
Kyoto Okayama Optical Low dispersion Spectrograph (KOOLS) with $R_\mathrm{C}$--band filter attached to OAO 1.88 m;
and Multiband Imager for Nayuta Telescope (MINT) with $R_\mathrm{C}$--band filter attached to NHAO 2 m. Two sets of
MITSuME systems at OAO 0.5 m and IAO 1.05 m are identically designed for monitoring transient objects such as gamma ray burst afterglows,
sharing nearly the same sky field at three wavelengths using two dichroic mirrors. The combinations of these telescopes
and instruments cover 26\arcmin $\times$26\arcmin\ field-of-view (FOV) with 1.53\arcsec\ pixel resolution at
OAO 0.5 m \citep{Kotani2005}, 5.0\arcmin $\times$ 4.4\arcmin\ FOV with 0.33\arcsec\ pixel resolution at OAO 1.88 m \citep{Yoshida2005},
12\arcmin $\times$ 12\arcmin\ FOV with 0.72\arcsec\ pixel resolution at IAO 1.05 m, and 11\arcmin $\times$ 11\arcmin\
FOV with 0.32\arcsec\ pixel resolution at NHAO 2 m. In addition, we made optical and near-infrared polarimetric
observations during two nights on UT 2014 December 24--27 by using the Nayoro Observatory 1.6 m Pirka telescope
of the Faculty of Science, the Hokkaido University (NO 1.6 m) and the Higashi--Hiroshima Observatory (HHO) Kanata 1.5 m
Optical and Near-Infrared telescope of Hiroshima Astrophysical Science Center, Hiroshima University (HHO 1.5 m).
We used a visible multi-spectral imager (MSI) with a polarimetric module  and
$R_\mathrm{C}$--band filter at NO 1.6 m \citep{Watanabe2012} and Hiroshima Optical and Near-Infrared camera (HONIR) with $R_\mathrm{C}$ and
$J$-band filters for HHO 1.5 m \citep{Akitaya2014}. In the imaging mode, NO 1.6 m/MSI has 3.3\arcmin $\times$ 3.3\arcmin\ FOV
with 0.39\arcsec\ pixel resolution, whereas HHO 1.5 m/HONIR has 10\arcmin $\times$ 10\arcmin\ FOV with 0.29\arcsec\ pixel resolution.
In the polarimetric mode, which is designed to use a focal mask for polarimetry, a Wollaston prism, and a half-wave plate, the NO 1.6 m/MSI
FOV is subdivided into two adjacent sky areas each having 3.3\arcmin $\times$ 0.7\arcmin\ FOV, and the HHO 1.5 m/HONIR FOV is
subdivided into five adjacent areas each having 9.7\arcmin $\times$ 0.75\arcmin\ FOV \citep{Watanabe2012,Akitaya2014}

The observed data were analyzed with standard techniques for CCD images.
Raw data were reduced by using flat field images taken with uniform screens on telescope domes and
with dark or bias obtained before or after the comet exposures. For imaging data, we combined individual
exposures into nightly composite images for each filter, excluding cosmic rays, background stars and galaxies
by using the same technique as that of \citet{Ishiguro2008}. $R_\mathrm{C}$--band flux calibration was performed in comparison
with field stars in the third U.S. Naval Observatory (USNO) CCD Astrograph Catalog (UCAC3), which ensured photometric accuracy of $\sim$0.1 mag \citep{Zacharias2009}.
We measured the instrumental magnitudes of all stars in the FOV and compared them with the catalog magnitudes to determine
the zero magnitudes of each $R_\mathrm{C}$--band image. 

For analysis of the polarimetric data, we followed the technique written in \citet{Kuroda2015}.
Because of a low signal to noise ratio (S/N), we did not analyze near-infrared polarimetric data.
Raw data were preprocessed by using flat and dark frames in the same manner as that for the imaging data.
We extracted source fluxes on ordinary and extraordinary parts of images by applying an aperture photometry technique.
The obtained fluxes were used for deriving the Stokes parameters normalized by the intensity,  $Q/I$
and $U/I$. The linear polarization degree ($P$) and the position angle of polarization ($\theta_\mathrm{P}$)
were derived by the following equations \citep{Tinbergen1996}:

\begin{eqnarray}
P = \sqrt{\left(\frac{Q}{I}\right)^2+\left(\frac{U}{I}\right)^2}~,
\label{eq:P}
\end{eqnarray}

\noindent and

\begin{eqnarray}
\theta_\mathrm{P}=\frac{1}{2}\tan^{-1}\left(\frac{U}{Q}\right)~.
\label{eq:theta}
\end{eqnarray}

We then derived the linear polarization degree commonly used for solar system objects ($P_\mathrm{r}$)
and the position angle of the polarization plane ($\theta_\mathrm{r}$) referred to the scattering plane,
which are given by

\begin{eqnarray}
P_\mathrm{r}=P \cos\left(2\theta_\mathrm{r}\right)~,
\label{eq:Pr}
\end{eqnarray}

\noindent and

\begin{eqnarray}
\theta_\mathrm{r}=\theta_\mathrm{P}-\left(\phi\pm90\arcdeg\right)~,
\label{eq:thetar}
\end{eqnarray}

\noindent where $\phi$ denotes the position angle of the scattering plane projected on the sky.
The sign in the parenthesis was chosen to meet the condition $0\leq (\phi \pm 90\arcdeg) \leq 180\arcdeg$.
The position angle of the polarized light from comets is generally perpendicular to the scattering plane
at the solar phase angle (Sun--object--observer's angle) $\alpha \gtrsim$30\arcdeg; thus, as expected,
$\theta_\mathrm{r}$ was $\sim$0\arcdeg.


\section{RESULTS}
\label{sec:results}

\subsection{Overall Appearance}
\begin{figure}
 \epsscale{1.0}
   \plotone{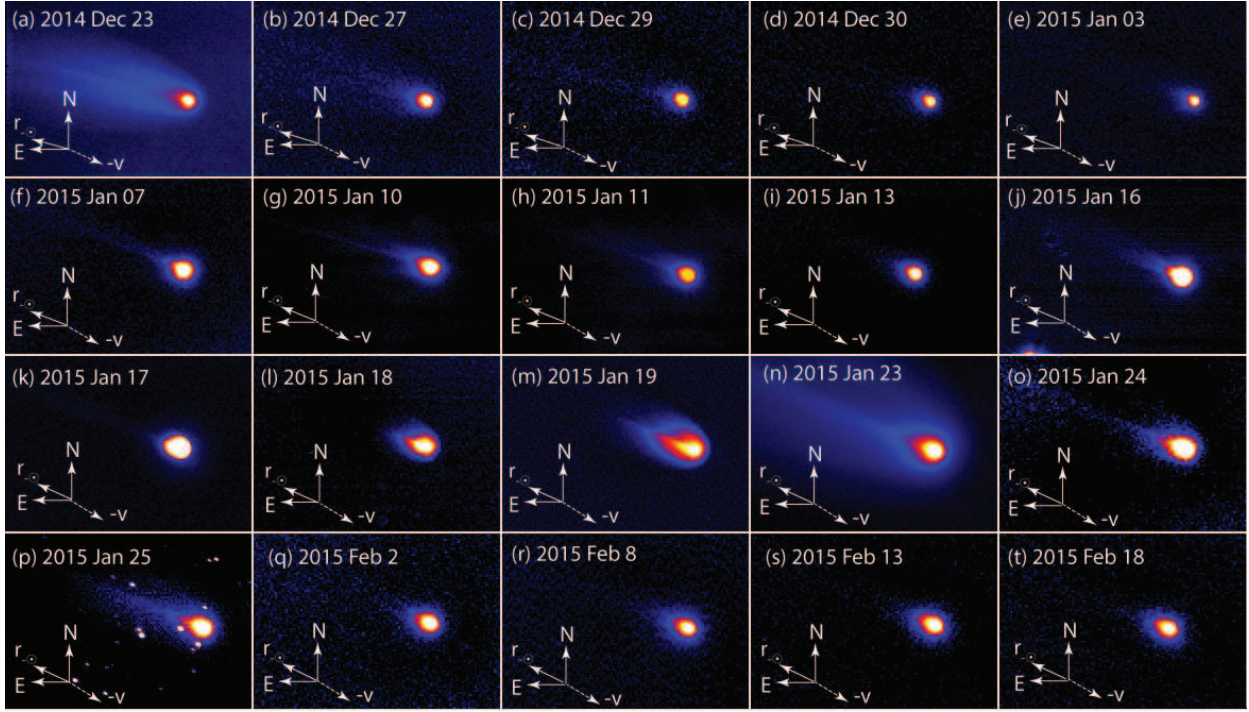}
  \caption{Time-series false color images of 15P taken with $R_\mathrm{C}$--band (wavelength 0.64 \micron)
  from (a) UT 2014 December 23 to (f) UT 2015 February 18. 
The FOV of each panel is 11.6\arcmin $\times$ 8.0\arcmin. All images have standard orientation
in the sky, that is, north is up and east is to the left. The anti-solar vectors ($r_{-\odot}$) and the negative
heliocentric velocity vectors ($-v$) are indicated by arrows. A dozen point--like sources appeared in (p) were not
erased by a star subtraction technique because of the short duration of exposures.}
\label{fig:TimeSeries}
\end{figure}

Figure \ref{fig:TimeSeries} shows time-series false-color composite images at the $R_\mathrm{C}$--band. We chose these
images because the $R_\mathrm{C}$--band is the most sensitive to cometary dust among the available filters. In fact, we examined
the contribution of the spherical gas component by using the same technique as that described in Section 3.1 of \citet{Ishiguro2014}. We found that
gas intensity took up only 10 $\pm$ 2\% (within an aperture at $\rho=10^4$ km from the nucleus) of $R_\mathrm{C}$--band total intensity
on UT 2014 December 26, which supports the weak gas flux contribution shown in Figure \ref{fig:TimeSeries}. Some images taken in close time intervals are
similar and are thus not shown separately. 
In the first image captured on UT 2014 December 23 (Figure \ref{fig:TimeSeries} (a)),  an envelope structure extends approximately toward the anti-solar
direction. It is likely that the envelope is related to the first outburst around UT 2014 December 16. The surface brightness of the envelope
reduced quickly and became undetectable after around UT 2014 December 29 (Figure \ref{fig:TimeSeries} (c)).
After that time, a near nuclear dust coma and dust tail remained. The coma and tail are attributed to steady activity of the comet
because both the shape and magnitude were almost constant over several months.
A dramatic change 
was observed in images after UT 2015 January 16. The inner coma brightened on UT 2015 January 16 (Figure \ref{fig:TimeSeries} (j)),
and dust ejecta appeared soon afterward  were stretched toward the anti-solar direction (Figure \ref{fig:TimeSeries} (k)--(n)).
The appearance of the dust cloud on UT 2015 January 23
(Figure \ref{fig:TimeSeries} (n)), is similar to the image taken on UT 2014 December 23 (Figure \ref{fig:TimeSeries} (a)), in which the comet was
enclosed by a widely expanded envelope. The envelope dimmed quickly, leaving behind a near-nuclear dust cloud similar to that from the pre-second
outburst. In summary, we observed at least two outburst ejecta superposed on the continuous activity of the comet during its
perihelion passage.

\subsection{Photometric Results}
\label{subsec:results}
To clarify the time variation of the activity in more quantitative manner, we conducted aperture photometry of the dust particles
in the inner coma. We set a constant physical aperture distance from the nucleus at $\rho=10^4$ km and integrated the signal
within $\rho$ by using the IRAF/APPHOT package. The aperture distances correspond to 7.4\arcsec--9.9\arcsec\ on the sky plane, which
is large enough to enclose the seeing disk sizes of these data (typically 2--3\arcsec) but small enough to detect daily changes in
dust production (an ejection speed of $\sim$100 m s$^{-1}$ was assumed).   In general, the measured magnitudes are determined
not only by the time-variable activity of comets but also by the observing geometry (i.e., distances and viewing angles).
To correct the latter effect, we converted the observed magnitudes into absolute values, which are magnitudes at a unit heliocentric
distance $r_\mathrm{h}$ = 1 AU, observer's distance $\Delta$= 1 AU, and solar phase angle $\alpha$ = 0\arcdeg.  These values were determined by 

\begin{eqnarray}
m_\mathrm{R}(1,1,0)=m_\mathrm{R} - 5~\log_{10}(r_\mathrm{h} \Delta) - 2.5\log_{10} \Phi(\alpha)~,
\label{eq:abs-mag}
\end{eqnarray}

\noindent where $m_\mathrm{R}$ and $m_\mathrm{R}(1,1,0)$ denote the observed and absolute magnitudes in the $R_\mathrm{C}$--band.
The third term in the right-hand side is given to correct the phase darkening,  which is given by

\begin{eqnarray}
2.5\log_{10} \Phi(\alpha) = b \alpha~,
\label{eq:phase-func}
\end{eqnarray}

\noindent where the constant $b$ was assumed to a generally quoted value for cometary dust particles, 0.035 mag deg$^{-1}$ \citep{Lamy2004}.
Over our observation period, the observed magnitudes were subtracted by 2.2--3.3 mag by Eqs. (\ref{eq:abs-mag})--(\ref{eq:phase-func})
to convert the absolute magnitudes.

\begin{figure}
 \epsscale{0.85}
   \plotone{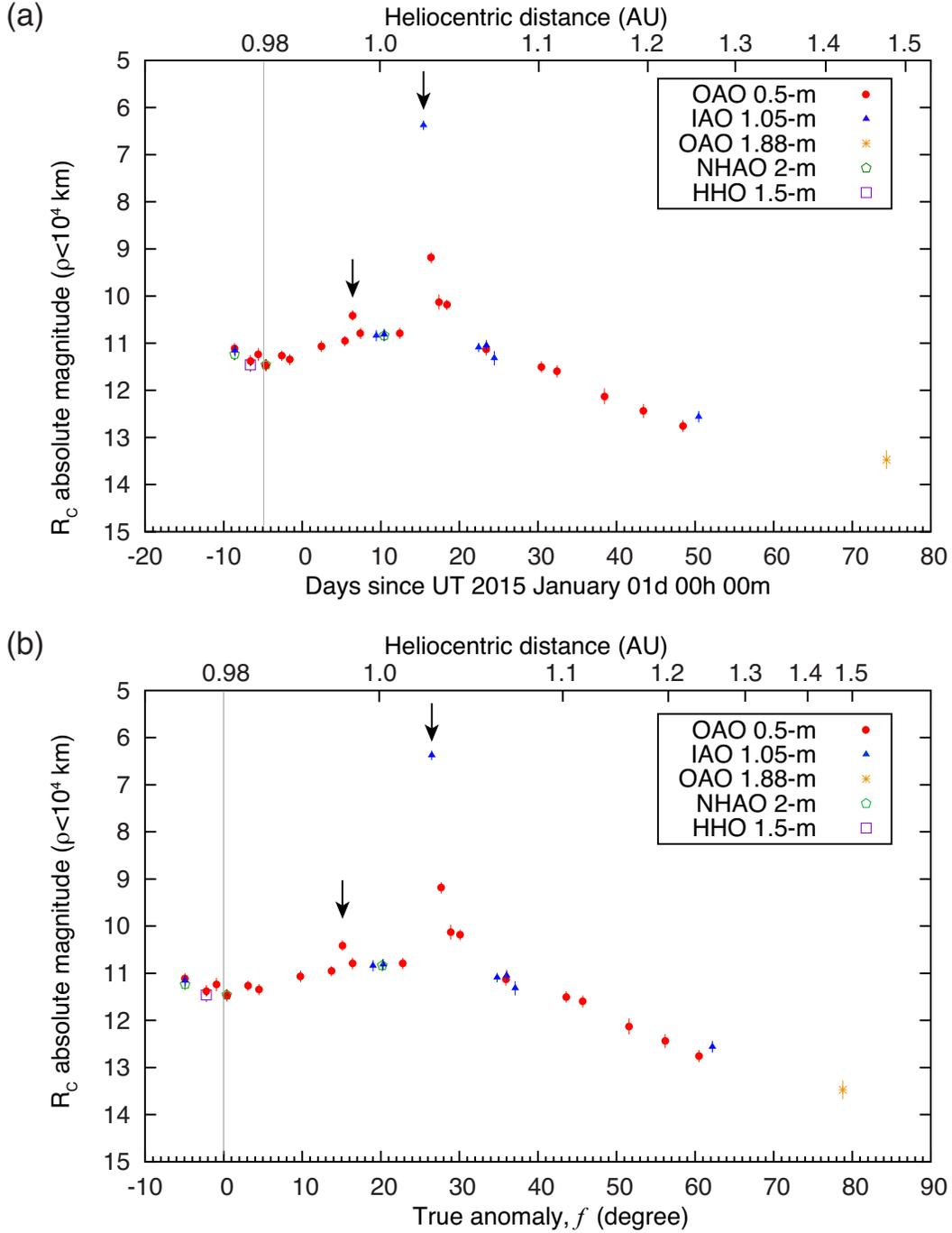}
  \caption{Aperture photometry magnitudes of dust coma at the physical distance $\rho<$10$^4$ km from the nuclear position
  with respect to (a) days since UT 2015 January 01 and (b) true anomaly. The corresponding heliocentric distances are
  shown at the tops of each graph. Thin vertical lines denote the perihelion. The brightening events on UT 2015 January 7
  and January 15 are indicated by downward arrows.}
  \label{fig:Mag}
\end{figure}

Figure \ref{fig:Mag} shows the absolute magnitudes with respect to (a) the observed time and (b) the true anomaly $f$; by definition,
the perihelion and aphelion occur at $f$ = 0\arcdeg\ and 180\arcdeg, respectively. The magnitude was almost constant or slightly decreased
by 0.03 mag day$^{-1}$ until the day of second outburst. A minor eruption appears to have occurred on UT 2015 January 7 with a true
anomaly of $f$ = 15.1\arcdeg, brightening the comet by 0.46 $\pm$ 0.14 mag.  A careful review of the above time-sequence images  revealed
a sharp tail on UT 2015 January 7--11, as shown in Figure \ref{fig:TimeSeries} (f)--(h).
Because it extended toward the position angle of 158.4 $\pm$ 1.0\arcdeg, which matches to anti-solar direction on that day (158.9\arcdeg),
we considered that a sudden ejection of fresh, small grains or ionized particles might have been stretched to the direction
soon after the minor eruption on UT 2015 January 7.

A more outstanding brightening was observed on UT 2015 January 16 ($f$ = 26.5\arcdeg). The
near-nuclear absolute magnitude was $m_\mathrm{R}(1,1,0)$ = 6.37 $\pm$ 0.10. After the brightening, the magnitudes
remained smaller than the pre-outburst magnitudes over three days and returned to a normal magnitude within about one week.
Although the image on UT 2015 January 23 (Figure \ref{fig:TimeSeries} (n)) showed the widespread envelope, it is likely that a large
part of the dust grains expanded beyond the aperture size of our photometry (i.e., $\rho>10^4$ km) and became obscured by a steady stream of ejecta
from the nucleus. To eliminate the magnitude excess on UT 2015 January 23, the outburst ejecta should have an ejection speed
able to reach the aperture (i.e., $\rho>10^4$ km). We determined that a large fraction of outburst ejecta should have an effective
speed of $\gtrsim$15 m s$^{-1}$ to reach the aperture radius within one week.

\subsection{Polarimetric Results}
\begin{figure}
 \epsscale{0.85}
   \plotone{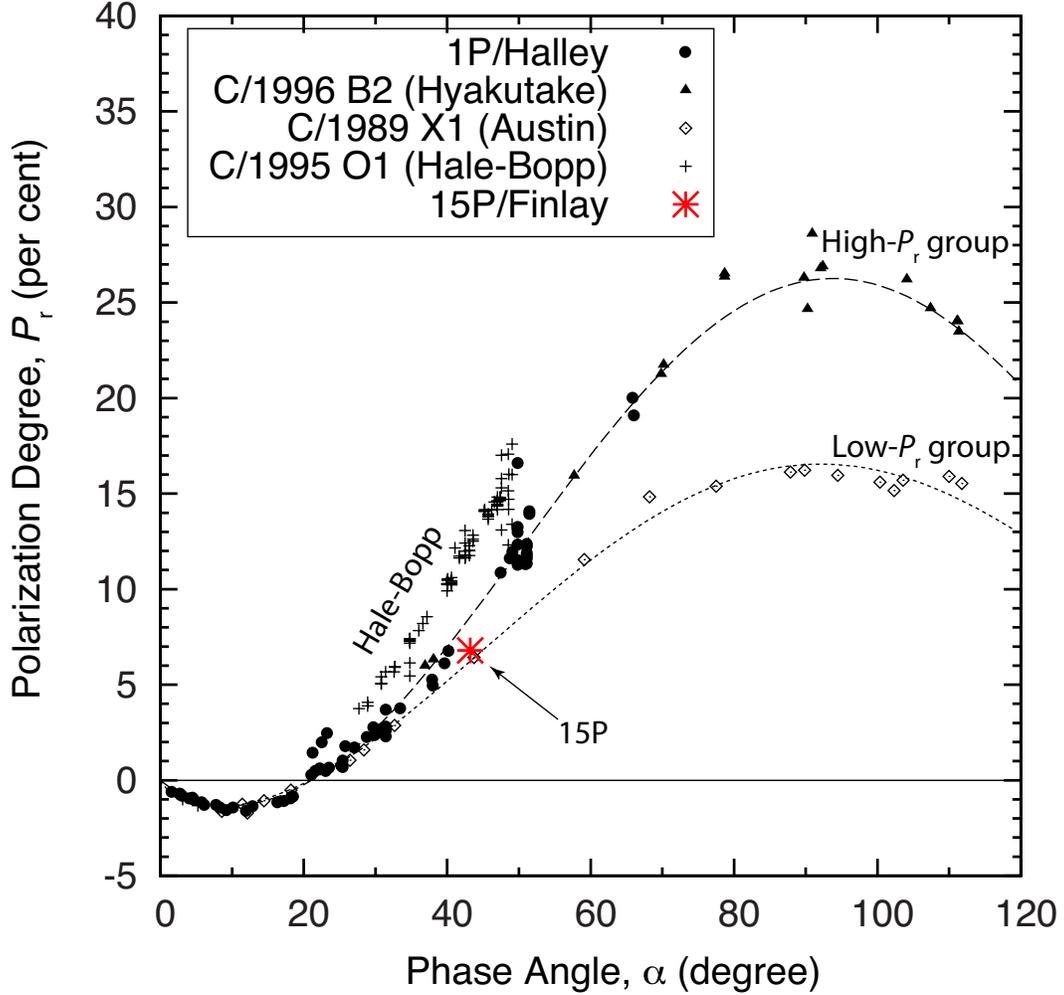}
  \caption{Phase angle dependence of $R_\mathrm{C}$--band polarization degree.
  For comparison, we show data of 1P/Halley and C/1996 B2 (Hyakutake)  as representatives
  of high-polarization comets and C/1989 X1 (Austin) as a representative of a low-polarization
  comet. These two classes of data were fitted by a trigonometric function.
  In addition, data of C/1995 O1 (Hale--Bopp) are shown, which exhibited an unusually
  high degree of polarization.
  The polarization data of other comets were acquired from NASA Planetary Data System "Database of Comet Polarimetry" \citep{Kiselev2010}.
  }
         \label{fig:pol}
\end{figure}

Polarimetric observation was conducted to find the optical similarities and differences between normal cometary dust and
the outburst ejecta. The polarimetric data were acquired about 10 days after the first outburst by using NO 1.6 m and HHO 1.5 m telescopes.
$R_\mathrm{C}$--band filters
were employed for our observations because this band is less contaminated by gaseous components and
is sensitive to the solar-like dust spectrum, as we mentioned above (Section \ref{subsec:results}). We integrated the signals of the observed data within apertures
of $\rho$ = 8,000 km and 12,000 km, and derived the polarization degree ($P_\mathrm{r}$) and the position angle of
the polarization vector ($\theta_\mathrm{P}$).
We obtained $P_\mathrm{r}$ = 6.5 $\pm$ 0.3\% on UT 2014 December 26 and $P_\mathrm{r}$ = 7.0 $\pm$ 0.5\%
on UT 2014 December 27.  The position angle was aligned to the normal vector of the scattering
plan, that is, $\theta_\mathrm{P}$ = 1 $\pm$ 4\arcdeg. No significant difference was noted within the errors
in $P_\mathrm{r}$ between measurements with the large and small apertures (i.e., $\rho$ = 8,000 km and 12,000 km).
Figure \ref{fig:pol} compares the polarimetric results with those of the other comets.
The phase angle dependence of the polarization degree is known to be classified into two groups: dust-rich comets with high polarization degrees
and gas-rich comets with low polarization degrees \citep{Levasseur1996}. Comet C/1995 O1 (Hale--Bopp) did not match these two categories
and showed a very high polarization degree \citep{Hadamcik1997}.
The polarization degree of 15P fell into common values of comets
between gas-rich and dust-rich groups. As described above, the surface brightness of the outburst ejecta faded out within $\sim$seven days from the
near-nuclear region in the case of the second outburst, which is likely similar to the case of the first outburst, and was undistinguishable from the background dust particles 
produced by continuous activity. Therefore, the similarity in polarization degree does not mean that the optical properties of
outburst ejecta are the same as those in normal comets. As a result of the data analysis, we learned that earlier follow-up observations within $\sim$three days are
required to determine the characteristics of such fresh particles during outbursts, such as fluffy or compact optical properties. This topic should be covered in future
observation planning.

\subsection{Order-of-Magnitude Estimates for the Outburst Ejecta Mass}
\label{subsec:order}
In this section, we provide an approximate but straightforward estimate of outburst ejecta mass to compare with previous outbursts,
and  subsequently present a more sophisticated estimate by conducting dynamical analysis of dust particles in Section  \ref{subsec:dynamics}.
We measured the total $R_\mathrm{C}$--band magnitudes with apertures large enough to enclose the entire visible dust cloud
for obtaining the total mass of the outburst materials. We set aperture radii of $\rho$ = 4.5\arcmin\ (UT 2014 December 23)
and 1\arcmin\ (UT 2015 January 16), and obtained $m_\mathrm{R}$ = 9.44 $\pm$ 0.10
(UT 2014 December 23) and $m_\mathrm{R}$ = 7.95 $\pm$ 0.10 (UT 2015 January 16). The magnitude
was $m_\mathrm{R}$ = 11.12 $\pm$ 0.12 on UT 2015 January 11--13, which was obtained just prior to the second outburst  and
well after the first outburst, when little outburst material was present as a result of radiation pressure sweeping.
By using Eqs. (\ref{eq:abs-mag})--(\ref{eq:phase-func}), we derived the absolute magnitudes of
$m_\mathrm{R}$(1,1,0) = 7.20 $\pm$ 0.10 (UT 2014 December 23), 8.82 $\pm$ 0.12 (UT 2015 January 11--13),
and 5.61 $\pm$ 0.10 (UT 2015 January 16). With these magnitudes, the optical cross-sections of the dust cloud
($C_\mathrm{c}$) were calculated by

\begin{eqnarray}
p_\mathrm{R}~C_\mathrm{c} = 2.24\times10^{22}~\pi~10^{0.4\left({m_\odot} - {m}_\mathrm{R}(1,1,0)\right)}~,
\label{eq:CrossSection}
\end{eqnarray}

\noindent where $p_\mathrm{R}$ is the geometric albedo ($p_\mathrm{R}$ = 0.04 was assumed), and $m_\odot$ = $-27.11$ is the
$R_\mathrm{C}$--band magnitude of the Sun at $r_\mathrm{h}$ = 1 AU \citep{Drilling2000}.  

By substituting $m_\mathrm{R}(1,1,0)$ in Eq. (\ref{eq:CrossSection}), we obtained $C_\mathrm{c}$ = 3.32 $\times$ 10$^{10}$ m$^2$
(UT 2014 December 23), 7.47 $\times$ 10$^{9}$ m$^2$ (UT 2015 January 11--13), and 1.44 $\times$ 10$^{11}$ m$^2$
(UT 2015 January 16). We attributed the cross-section on UT 2015 January 11-13 to dust grains ejected
via continual activity, which is irrelevant to these outbursts. We subtracted  this value from
the other two values and obtained the cross-section associated with the first outburst, $C_\mathrm{c}$ = 2.56 $\times$ 10$^{10}$ m$^2$,
and the second outburst $C_\mathrm{c}$ = 1.36 $\times$ 10$^{11}$ m$^2$.
There is a possibility that we miss some fraction of the cross-section from the first outburst because our data might not cover its entire dust cloud
on UT 2014 December 23. We thus considered the total cross-section of the first outburst ejecta
as  the lower limit (i.e., $C_\mathrm{c}$ $\ge$ 2.56 $\times$ 10$^{10}$ m$^2$ for the first outburst ejecta). 

The ejecta masses of the outbursts can be given by
\begin{eqnarray}
M_\mathrm{d}=\frac{4}{3}\rho_\mathrm{d}a_\mathrm{eff}C_\mathrm{c}~,
\label{eq:mass}
\end{eqnarray}

\noindent where $a_\mathrm{eff}$ and $\rho_\mathrm{d}$ are the effective dust grain radius and mass density, respectively.
The effective radius is given by

\begin{eqnarray}
a_\mathrm{eff}=\frac{\int^{a_\mathrm{max}}_{a_\mathrm{min}}a^{3-q} da}{\int^{a_\mathrm{max}}_{a_\mathrm{min}}a^{2-q} da}~,
\label{eq:aeff}
\end{eqnarray}

\noindent where $q$ denotes the power index of the size distribution in the range of $a_\mathrm{min}\leq a \leq a_\mathrm{max}$.

With assumptions of $\rho_\mathrm{d}$ = 1000 kg m$^{-3}$ and $a_\mathrm{eff}$ = 1 $\times$ 10$^{-6}$ m (i.e., 1\micron), $C_\mathrm{c}$ gives
$M_\mathrm{d}$ $>$ 3.41 $\times $10$^7$ kg for the first outburst and $M_\mathrm{d}$ = 1.81 $\times$ 10$^8$ kg for the second outburst. These values are
equivalent to masses of $R_\mathrm{c}>$ 20 m and 35 m bodies. It should be noted, however, that
this simple assumption likely underestimated the dust mass.  Although the small particle ($a_\mathrm{eff}$ = 1 \micron) assumption is occasionally quoted in
previous research, abundant evidence exists for large particles from comets, which is subsequently discussed. Thus, it is better to consider that the masses
should be the lowest limits. \citet{Ye2015} derived $C_\mathrm{c}$ = 7 $\times$ 10$^9$ m$^2$ for the first outburst
and $C_\mathrm{c}$ = 2 $\times$ 10$^{10}$ m$^2$ for the second outburst, which are smaller than our estimates.
In addition, they reported $M_\mathrm{d}$ = (2--3) $\times$ 10$^5$ kg for the first outburst and $M_\mathrm{d}$ = (4--5) $\times$ 10$^5$ kg
for the second outburst, which are more than two orders of magnitude less than our estimates. Although we were 
unable to verify their results with the information available in their paper, we found that these orders-of-magnitude estimates reveal that
the ejecta masses of 15P outbursts are equivalent to that of the 332P event (10$^8$--10$^9$ kg) \citep{Ishiguro2014}.

\subsection{Motion of Dust and Gas Ejecta in the Second Outburst Images}
\label{subsec:dustgas}
\begin{figure}
 \epsscale{0.5}
   \plotone{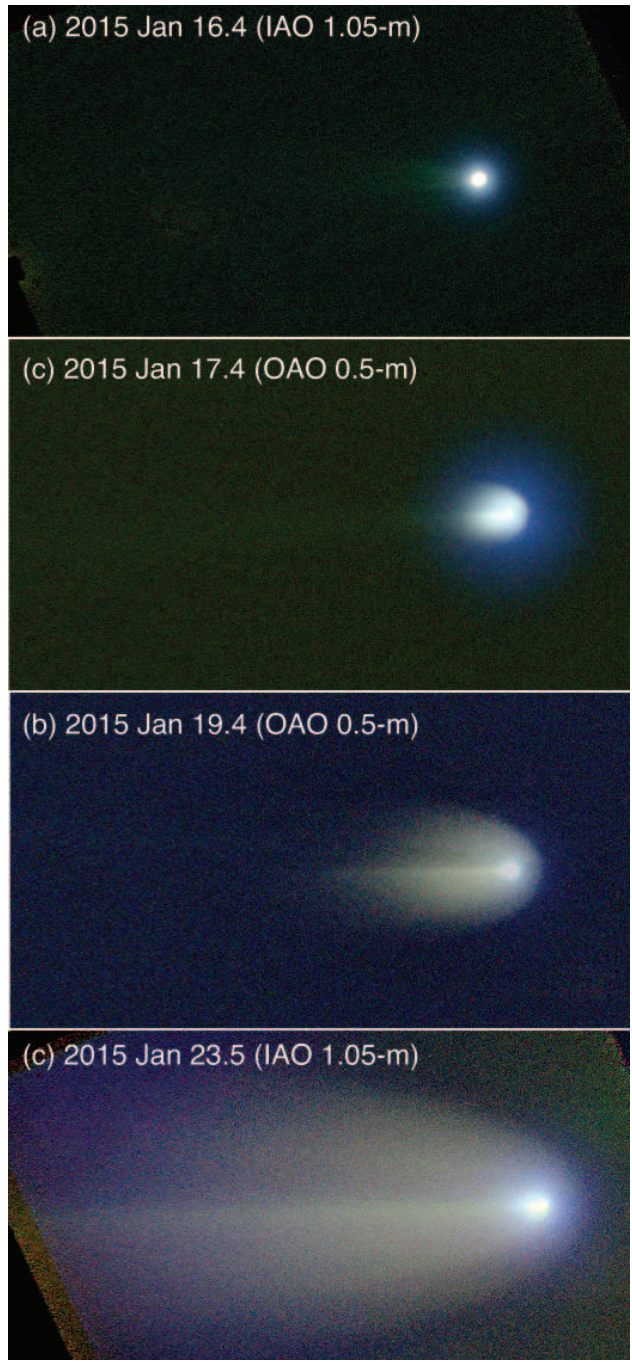}
  \caption{Selected images after the second outburst. We allocated {\sl g}$'$-band images as blue, $R_\mathrm{C}$--band images
  as green, and $I_\mathrm{C}$--band images as red to create these color images.
  These images are rotated to align the anti-solar vector ($r_{-\odot}$) 
  to the horizontal direction of each image. The FOV of each panel is 13\arcmin $\times$ 7\arcmin. }
    \label{fig:Col2nd}
\end{figure}

We further investigated the images soon after the second outburst. Figure \ref{fig:Col2nd} shows images taken after
UT 2015 January 16.4 on the day in which brightening in Figure \ref{fig:Mag} was detected. We do not show the image taken on UT 2015 January
18 because of its bad quality (the total exposure time was only 3 min due to unstable weather). To clarify the movement of
dust grains via solar radiation pressure, we rotated these images to match the Sun--comet's direction to the horizontal direction.
We produced color images assigning {\sl g}$'$-band images (the effective wavelength and
the full width at half-maximum of $\lambda_\mathrm{e}$ = 483 nm and  $\Delta \lambda$ = 134 nm) to blue; $R_\mathrm{C}$--band
images ($\lambda_\mathrm{e}$ = 655 nm and $\Delta \lambda$ = 121 nm) to green; and $I_\mathrm{C}$--band images
($\lambda_e$ = 799 nm and  $\Delta \lambda$ = 157 nm)  to red colors. In these images, a whitish component expanded in space
and simultaneously stretched toward the anti-solar direction. On the contrary hand, a bluish component expanded spherically with respect
to the nuclear position. In general, prominent emissions associated with C$_2$ and CN appear in the {\sl g}$'$ band, weak emissions with
NH$_2$ in the $R_\mathrm{C}$--band, and negligibly faint signals with NH$_2$ and CN in the $I_\mathrm{C}$--band
\citep{Brown1996,Meech2004}. In theory, dust particles are accelerated by solar radiation pressure, whereas
neutral gas molecules are less sensitive to the radiation pressure and expand almost spherically. Thus, it is reasonable to consider that the whitish
elongated structure originated from scattered sunlight from dust particles and that the bluish structure originated from CN and C$_2$ emissions.

We attempted to extract the signals from the second outburst and to discriminate the gas flux from the
scattered light by dust grains. The observed angles rotated very little over 2015 January 11--23 (that is,
0.4\arcdeg\ for the phase angle, 2.4\arcdeg\ for the position angles of the anti-solar vector, and 1.4\arcdeg\ for the position angle
of negative orbital velocity). Therefore, we subtracted the pre-outburst signals by using images observed on 2015 January 11 (IAO 1.05 m) and
2015 January 13 (OAO 0.5 m) from post-outburst images to adjust the apparent sizes of the comet based on the geocentric
distances without rotating these images. Next, assuming that $I_\mathrm{C}$--band images
have little gas contamination, we  subtracted them from {\sl g}$'$-band images by adjusting the $I_\mathrm{C}$--band intensity scales
to minimize the extended dust structure (i.e. dust tail) in the residual images to produce gas intensity maps.
Specifically, we detected a very weak spherical component in the $I_\mathrm{C}$--band images that likely originated
from weak CN emissions at 790--820 nm, which were noticeable until UT 2015 January 17 but were unclear after January 19.
We subtracted the faint spherical component in the $I_\mathrm{C}$--band images by using the above gas intensity maps, scaling the gas intensities
to obscure the spherical components in the residuals to produce new dust intensity maps assuming that the intensity
distributions of the gas components are the same in {\sl g}$'$ and $I_\mathrm{C}$--bands.

\begin{figure}
 \epsscale{0.8}
  \plotone{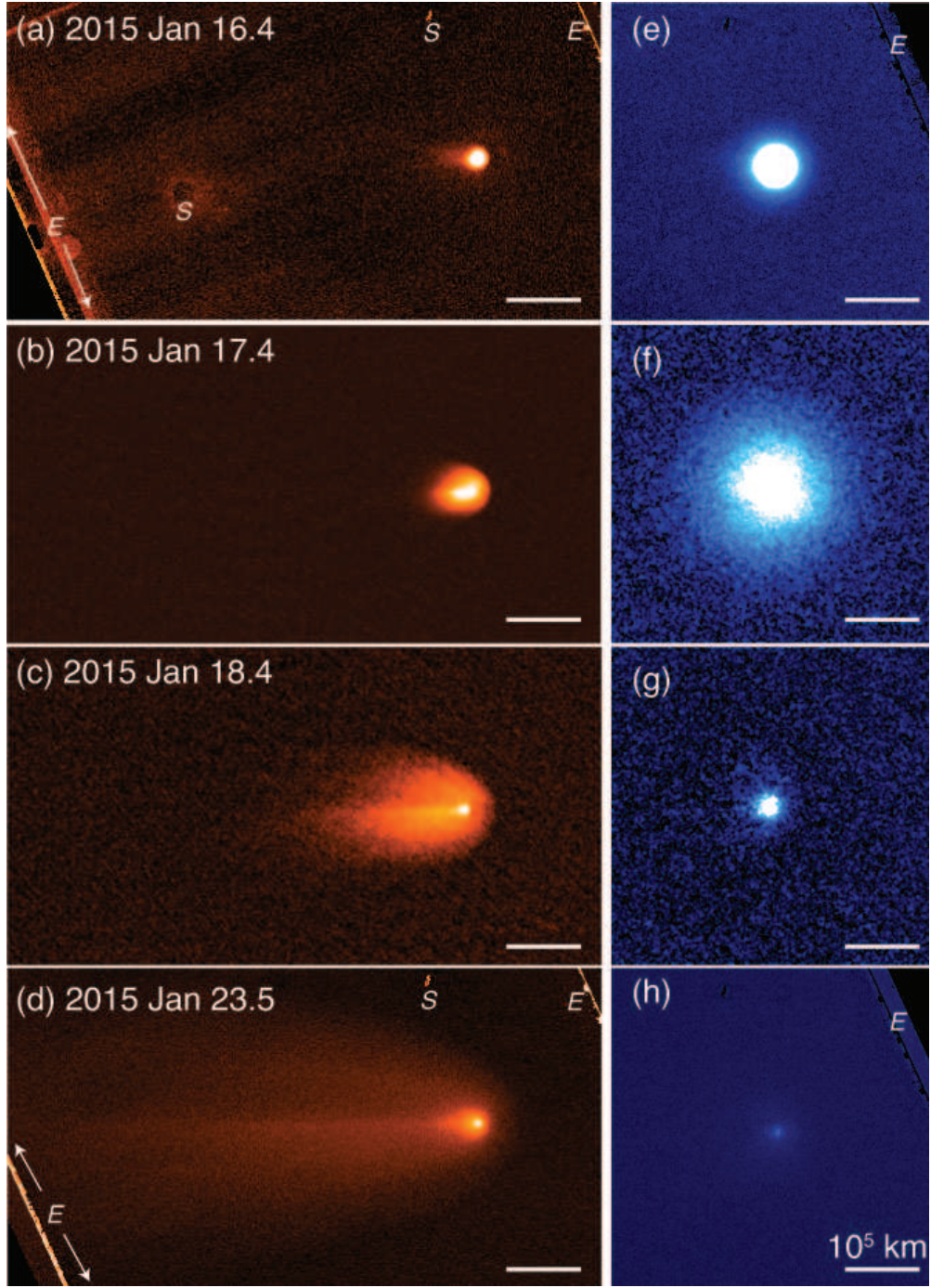}
 \caption{(a)--(d) Dust and (e)--(h) gas distributions derived by using the same images as those in Figure \ref{fig:Col2nd}. To
 view faint diffuse structures, we applied 3 $\times$ 3 pixel boxcar smoothing to these images.  The physical
   scale at the position of the comet, 10$^5$ km, is indicated by horizontal lines. Remnants of star subtraction and CCD artifacts
   associated with the edges of frames are shown by "$S$" and "$E$". The FOV of each panel is 13\arcmin $\times$ 7\arcmin\ for dust images
   and 7\arcmin $\times$ 7\arcmin\ for gas images.}
       \label{fig:gasdust}
\end{figure}

\begin{figure}
 \epsscale{0.8}
  \plotone{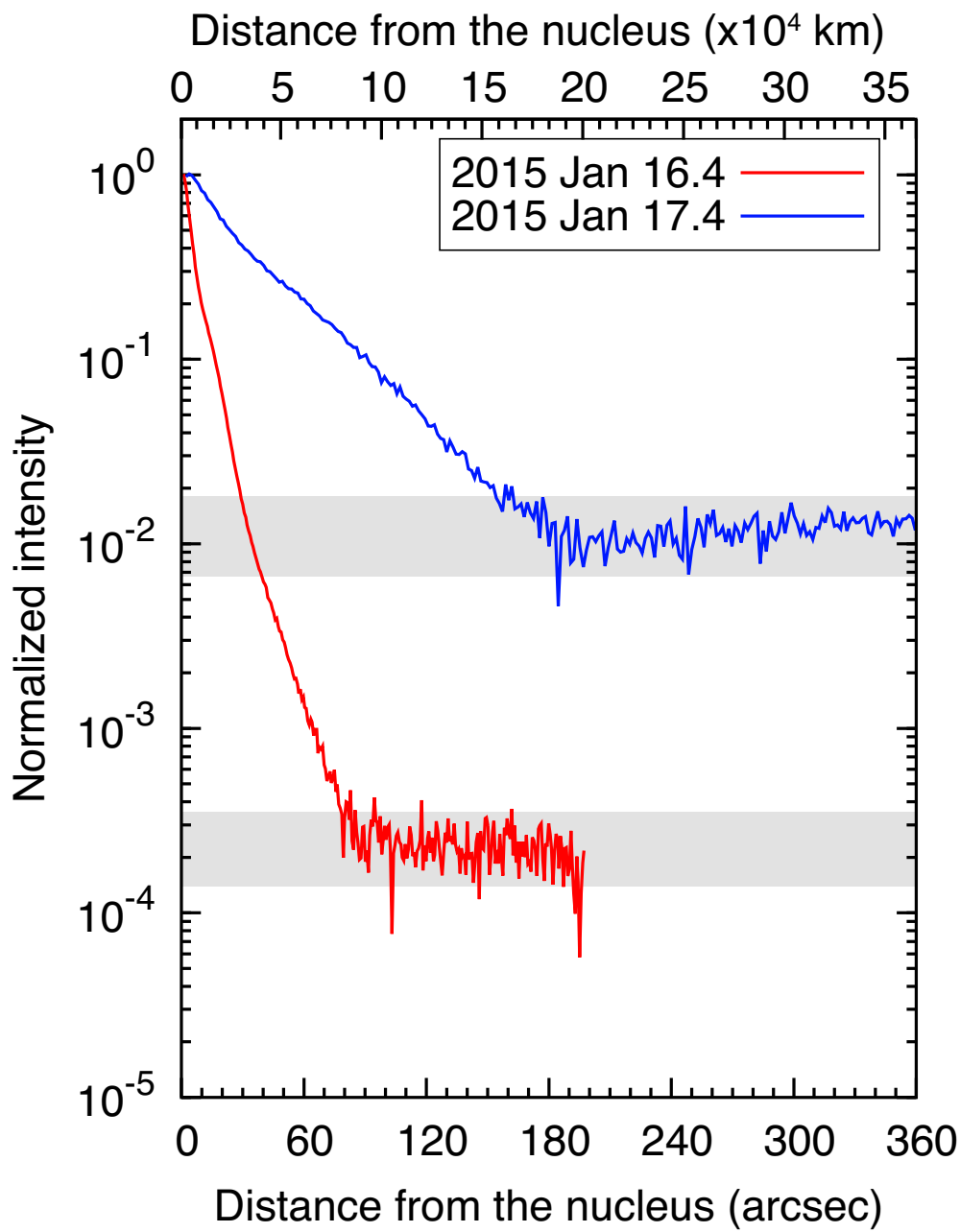}
 \caption{Radial surface intensity profiles of the gas component on UT 2015 January 16.4 and 17.4. We plotted the average
 	within an 1\arcsec\ annulus.  We adjusted the sky levels to 3 $\times$ $\sigma$ (the standard deviation of the sky
	background area, indicated by gray boxes)  to avoid fluctuation of the values to negative in the semilogarithmic plot.
	Apparent distance and physical length from the nucleus are shown at the bottom and top of the horizontal axes, respectively.}
       \label{fig:radial}
\end{figure}

\begin{figure}
 \epsscale{0.8}
  \plotone{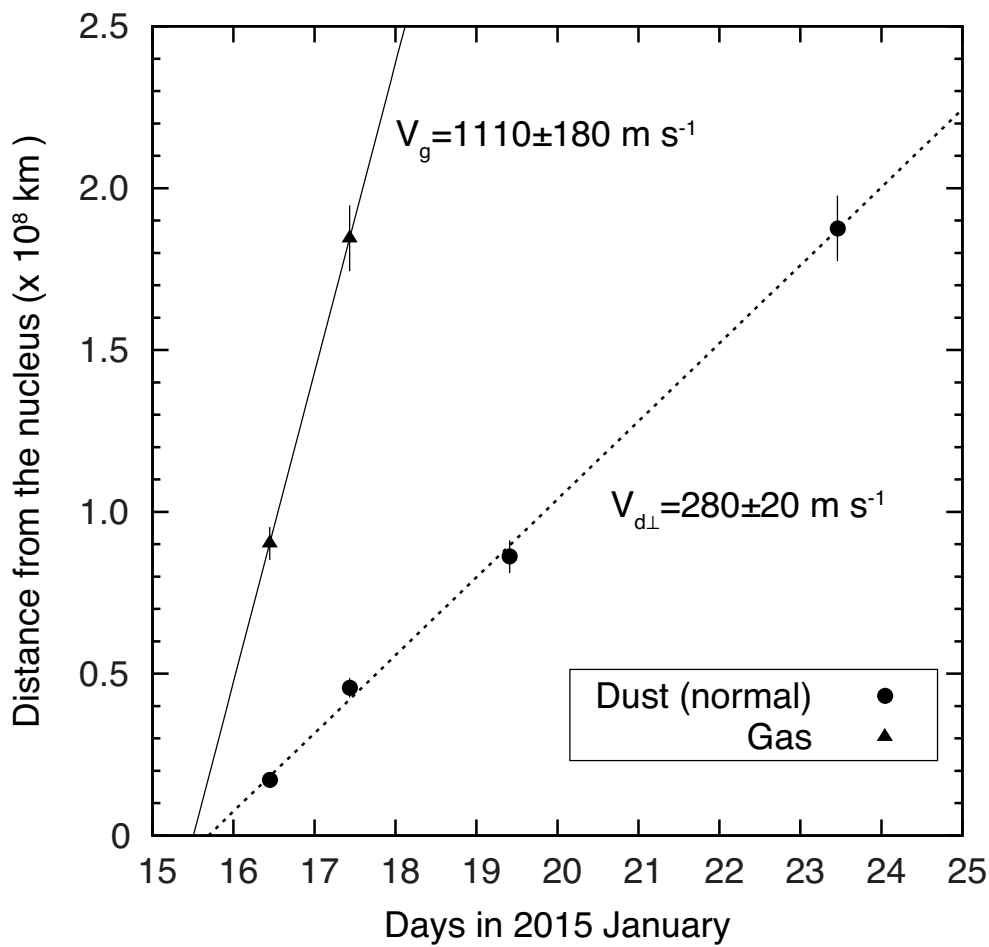}
 \caption{Distance from the nucleus to outer gas envelope (filled triangles) and half width of the dust envelope (filled circles). We determined
 the onset time of UT 2015 January 15.5 $\pm$ 0.2 from the gas envelope and UT 2015 January 15.7 $\pm$ 0.1 from the dust envelope. 
 The gas expansion speed was $V_\mathrm{g}$ = 1110 $\pm$ 180 m s$^{-1}$. The dust expansion speed with respect the projected
 anti-solar vector was $V_\mathrm{d\bot}$ = 280 $\pm$ 20 m s$^{-1}$.}
       \label{fig:speed}
\end{figure}

The resultant images are shown in Figure \ref{fig:gasdust}.
As expected, we extracted a spherical structure centered on the nucleus for the gas component associated with the second
outburst (Figure \ref{fig:gasdust} right). The surface brightness reduced rapidly in a few days not only because the gas 
diffused out in space, but also because the molecules had lifetimes equivalent to the observed period, i.e., two days for CN and five days for C$_2$
during the active solar phase (Huebner et al. 1992). 
We then examined the radius of the gas component from the images. Figure \ref{fig:radial} shows the radial profile of the gaseous coma.
Considering the sky background noise, we derived the radii of 89$\pm$5\arcsec\ on UT 2015 January 16.4
and 182$\pm$10\arcsec\ on UT 2015 January 17.4. Assuming that the gas expanded at a constant speed since its ejection,
we derived the onset time and expansion speed of the gas component as UT 2015 January 15.5 $\pm$ 0.2 and
$V_\mathrm{g}$ = 1,110 $\pm$ 180 m s$^{-1}$ (Figure \ref{fig:speed}). We quoted the maximum ranges of these parameters
as these errors in consideration of the margin of measurement because we had only two data points, which is insufficient
for deriving the errors by using the least squares method. We found that the derived gas speed was slightly faster than the observed
values of CN emission at other comets around 1 AU probably  because we derived the speed at a very large
distance from the nucleus ($\approx10^8$ km), where the gas flow velocity continued to increase \citep{Ip1989,Krankowsky1986}.

In the left-hand column of Figure \ref{fig:gasdust}, we show the dust component images. A short tail was visible on the first night (a),
and an irregular cloud appeared on the second night (b) that extended toward the anti-solar direction via the radiation pressure.
From the apparent cloud size, we derived the ejection onset time. We  focused on the widths of
the dust cloud perpendicular to the Sun--comet direction because dust particles were not accelerated by radiation pressure along that direction.
Figure \ref{fig:speed} shows the time evolution of the dust cloud width. The width increased linearly with the progression of time.
By using a linear least squares method, we obtained the onset time on UT 2015 January 15.7 $\pm$ 0.1 for dust particles. Although this result agrees with
the onset time of gas within the accuracy of our measurements, the trivial lag may suggest that the dust required more time to reach
terminal velocity, which should be accelerated by the gas outflow. We conclude that the second outburst occurred on UT 2015 January
15.6--15.7 from gas and dust components. In addition, we derived the dust speed perpendicular to the Sun--comet direction as $V_\mathrm{d\bot}$ = 280 $\pm$ 20 m s$^{-1}$.
Here, we should note that this speed might be misleading. Because we measured the speed perpendicular to the Sun--comet direction,
we must underestimate the ejection speed for dust particles in  $V_\mathrm{d\bot}$.

\subsection{Ejecta Dynamical Model}
\label{subsec:dynamics}
To derive the ejection speed, mass, and kinetic energy of the dust cloud in a comprehensive manner, we conducted a model simulation to reproduce the observed
morphologies of dust ejecta from the second outburst by following the scheme in \citet{Ishiguro2013}. 
The morphologies of dust clouds are generally determined by the ejection speeds and the effects of solar radiation pressure, which can be approximately given by
a function of grain size. Therefore, the sizes and ejection speeds of the dust particles can be determined essentially through an investigation of dust cloud morphologies.
For a spherical particle, the ratio of the radiation pressure with respect to the solar gravity ($\beta$) is given as $\beta$ = $0.57\times10^{-4}~Q_{pr}/{\rho_d}a$,
where $a$ and $\rho_d$ are dust radius (m) and mass density (kg m$^{-3}$), and $Q_{pr}$ is a radiation pressure coefficient \citep{Burns1979,Finson1968}.
Here, we assumed $\rho$ = 1$\times$10$^3$ kg m$^{-3}$ and $Q_{pr}$ = 1.
For convenience of fitting, we assumed that the dust ejecta consisted of two components,
a high--speed envelope and a low--speed tail. We set the ejection epoch of UT 2015 January 15.5,
which gives negligible influence in the result within the error range (i.e., 0.1 day). For simplicity, we assumed that dust particles were ejected symmetrically with respect
to the solar direction within a cone of a half opening angle of $w$. We employed a size--dependent ejection speed of $V_\mathrm{ej}$ = $V_0 a^k$
and power--law size frequency distribution given by $N(a)$ = $N_0 a^{-q}$ in the range between $a_\mathrm{min}$ and $a_\mathrm{max}$.
Because the ranges of these parameters and fitting scheme are same as those given in \citet{Ishiguro2013}, we do not describe the details here.

\begin{figure}
 \epsscale{1.00}
   \plotone{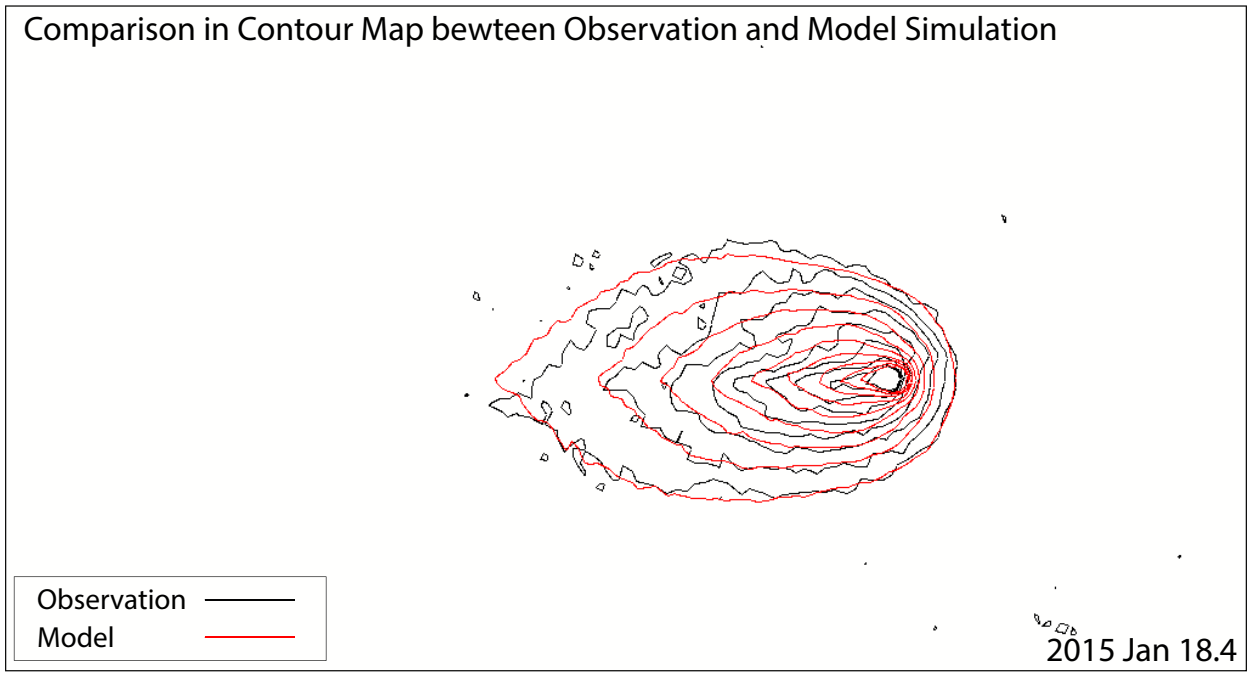}
  \caption{Comparison of the contour maps between observation and model simulation for the data on 2015 January 19. 
  Similar to that in Figure \ref{fig:gasdust}, we rotated the contour to match the anti-solar direction to the horizontal axis and
  trimmed the data in the region of the FOV of 13\arcmin$\times$7\arcmin.
  We applied box car smoothing to the simulation data to match the spatial resolution of the observed contour.}
         \label{fig:model}
\end{figure}

Through the fitting, we derived the best-fit parameters as shown in Table \ref{tab:parameter}.
Because large particles are not sensitive to solar radiation pressure and still resided near the nucleus, we found it impossible to derive the maximum size of
particles. We fixed the minimum value for $\beta$ to $6\times10^{-4}$ (i.e., $a=$1 mm) on the ground that large cometary boulders may not be ejected efficiently through  outbursts
\citep{Bertini2015}. Figure \ref{fig:model} shows a comparison of the observed dust cloud morphology and our best-fit model, which broadly
match. However, the entire morphology was not perfectly reconstructed by our model probably because the dust ejection was not symmetric
with respect to the solar direction. Although we do not intend to upgrade the model fitting by fine tuning the central axis of the dust emission here,
we would insist that the outburst location might deviate slightly from the sub-solar point because of this asymmetry. 
The maximum $\beta$ was determined well to be $\beta_\mathrm{max}$ = 1.6 $\pm$ 0.2  because our observation data covered the faint end of the dust cloud,
where particles with $\beta_\mathrm{max}$ were dominant. 
To fit the width and sunward extension of the cloud, we obtained the best fit parameters, $V_\mathrm{0}$ = 450 $\pm$ 30 m s$^{-1}$ for the envelope
particles and $V_\mathrm{0}$ = 330$^{+60}_{-30}$ m s$^{-1}$ for the tail. These parameters resulted in the maximum speed for the smallest particles
of $V_\mathrm{max}$ = 570 $\pm$ 40 m s$^{-1}$. For confirmation, we calculated the speed perpendicular to the Sun--comet direction with these simulation results,
$V_\mathrm{d\bot}$ = $V_\mathrm{max} ~\sin \left(w\right)$ = 285 $\pm$ 40 m s$^{-1}$, which is consistent with the value given in Section \ref{subsec:dustgas} 
(i.e., $V_\mathrm{d\bot}$ = 280 $\pm$ 20 m s$^{-1}$). 
From the model fitting, we determined that $\sim$20\% of the cross-section originated from the envelop, whereas $\sim$80\% was from the tail and coma.
We calculated the total mass and kinetic energy for the second outburst by using the parameters derived from the simulation (see Table \ref{tab:parameter})
and obtained $M_\mathrm{d}$ = 7.0 $\times$ 10$^8$ kg and $E_\mathrm{d}$ = 6.5 $\times$ 10$^{12}$ J for the second outburst.

\section{DISCUSSION}
\label{sec:discussion}

\subsection{Maximum Radiation Pressure Force and Constraint on Dust Physical Properties, $\beta_\mathrm{max}$}
Cometary outbursts provide opportunities to investigate fresh cometary materials that are embedded in the surface processed layers
but appear when a large amount of materials is ejected via explosions. In particular, the 15P event in 2015 January  provided a unique
opportunity to determine the maximum $\beta$ value with good precision for several reasons.
Because we determined the precise onset time of the outburst, it was easy to follow the motion with respect to the reference time. In addition, the
outburst ejecta were observed at the large phase angle of $\alpha$ = 45\arcdeg\ (c.f. $\alpha$ = 19\arcdeg\ for 332P and 17\arcdeg\ for 17P), which enabled 
observation of the dust motion toward the anti-solar direction more clearly. Moreover, the outburst occurred at a smaller heliocentric distance (1.0 AU),
where the acceleration by radiation pressure was detected more effectively  than those at distant locations such as 1.6 AU for 332P and 2.4 AU for 17P.

It is well known that radiation pressure force depends on not only size but also and composition.
The $\beta$ values have been investigated theoretically considering fluffy dust particles \citep{Mukai1992} and a variety of compositions \citep{Burns1979}.
In section \ref{subsec:dynamics}, we adopted a simple model that $\beta$ is inversely proportional to the size under an assumption of a radiation pressure
coefficient of $Q_\mathrm{pr}$ = 1.
However, this assumption holds only when the particle size is larger than the optical wavelength ($\lambda$ = 0.5 \micron). $Q_\mathrm{pr}$ has
almost constant value for $a>$0.1--0.3 \micron\ and significantly drops as the size of dust grains decreases \citep{Ishiguro2007}. 
For fluffy dust particles, $\beta$ is less dependent on the aggregate size but similar to that of each constituent \citep{Mukai1992}.
\citet{Wilck1996} calculated the $\beta$ values of dust particles by using a core--mantle spherical particle model
composed of silicate and amorphous carbon with different porosities and suggested that cometary dust particles have
$\beta_\mathrm{max}$=1--1.8.  Transparent materials such as silicates and water ice tend to have small $\beta$ maximum values (i.e.,
$\beta_\mathrm{max}<$1), whereas that of absorbing particles such as carbon is $\beta_\mathrm{max}>$1 \citep[see e.g.,][]{Kimura2016}. The $\beta_\mathrm{max}$ value determined in our measurement, 1.6 $\pm$ 0.2, is consistent with that in the porous absorbing particles model \citep{Wilck1996} and silicate-core, organic-coated grains model for dust
aggregates \citep{Kimura2003}. However, our value is inconsistent with those of silicate spheres ($\beta_\mathrm{max}<$1) and organic spheres ($\beta_\mathrm{max}>$3).

\subsection{Outburst Frequency}
Cometary outbursts have been observed in a wide variety of comets.
There are some references that analyzed historical observation of cometary outburst \citep[e.g.][]{Hughes1990}.
However, the occurrence frequency and 
mass production rate have not been studied well because there is no coordinated observation system for monitoring comet magnitude.
Some outbursts could have been missed because the magnitude contrast was too weak before and after to be noticed as an outburst or
because the observation condition was not suitable for detection (i.e., too close to the Sun or Moon). Nevertheless,
some outburst events have been reported in a voluntary manner by amateur observers. For example, the 17P event was first reported by
A. Henriquez Santana (Spain), who noted sudden brightening to 8.4 mag by using a 0.2-m reflector \citep{Buzzi2007}. 332P was discovered in Japan during its outburst
by S. Murakami and K. Ikeya. They observed this event by looking into the eyepieces of 0.46-m and 0.25-m telescopes when it reached $\sim$9 mag 
\citep{Murakami2010}. Moreover, 15P double outbursts with brightening to 8--9.4 mag were also reported by amateurs through a mailing list of comet observers
\citep{Ye2015}.

It has been suggested that 17P and 332P events were caused by a phase change of amorphous water ice \citep{Li2011,Ishiguro2013}, although
other mechanisms such as rotational breakup of brittle cometary nuclei cannot be ruled out \citep{Li2015}. Excavations via impacts
might create an environments favorable to outbursts, as suggested in \citet{Beech2002}.
Because the energy per unit mass is similar
to both events, we conjecture that 15P events were also caused by a phase change of amorphous water ice.
Here, we consider the frequency of such outbursts in the inner ($r_\odot\lesssim5$ AU) solar system, where the physical mechanism of general cometary activity is confined to sublimation of water ice rather than that of super volatiles such as CO  \citep{Jewitt2009}.
For the reason, we excluded frequent outbursts at 29P/Schwassmann-Wachmann 1 beyond the Jupiter orbit.
Moreover, we restricted our discussion to JFCs and Encke-type comets (ETCs) because significant observation samples are not available for Halley-type comets
(HTCs) and long-period comets (LPCs).
We applied a sample cut of the perihelion distance $q\lesssim5$ AU to the JPL Small-Body Database Search Engine\footnote{http://ssd.jpl.nasa.gov/sbdb_query.cgi} and retrieved 505 JFCs and ETCs in the list of known comets as of the end of 2015. We regarded disintegrated comets as a single object (e.g. 73P/Schwassmann-Wachmann 3 B, C, G... constituted a united body). Moreover, we excluded disappeared or dead comets (3D/Biela and 5D/Brorsen), designated as "D/," and main-belt comets, which are 
unlikely to contain amorphous ice \citep{Prialnik2009}.
Among JFCs and ETCs, 357 objects passed their perihelion at least once between 2007 October and 2015 December, which make up
$\sim$70\% of the entire population. This means that $\sim$70\% of JFCs and ETCs might have a chance to display outburst activities
owing to the additional heat from the sun near their perihelia.

We attempted to find outburst events through the Smithsonian Astrophysical Observatory/National Aeronautics and Space Administration (SAO/NASA) Astrophysics Data System (ADS) Astronomy Query Form by inputting "comet" and "outburst" as abstract terms and manually plumbing the results by reading the abstracts.
In addition, we added one event at 205P which showed an outburst showing similar appearance to 15P (Seiichi Yoshida, private communication).
We found that 15 outburst events occurred at 11 comets since 2007 (Table 4).  We chose the arbitrary period beginning in 2007
because researchers have increased their consciousness toward outbursts since then, as motivated by the 17P event.

\begin{figure}
 \epsscale{0.85}
   \plotone{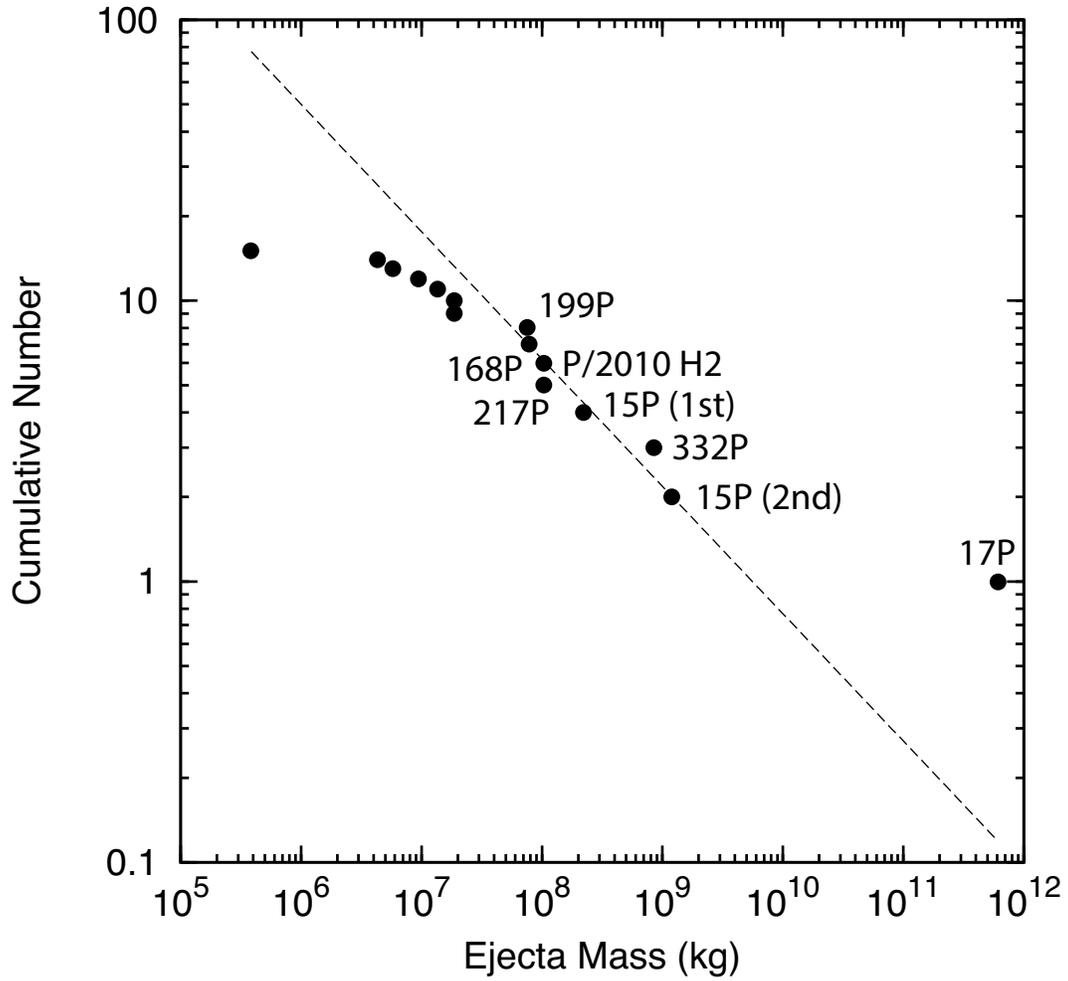}
  \caption{Cumulative distribution of outburst ejecta mass observed since 2007. We assumed that outburst ejecta of these comets (except 17P, 332P and 217P) have the same size distribution as 15P.}
         \label{fig:cumM}
\end{figure}

Table 4 shows some minor outbursts detected by observers with larger telescopes because these
comets drew their attentions due to space mission targets (81P/Wild 2 and 103P/Hartley 2) and repeated outbursts
at 17P and 15P. Figure \ref{fig:cumM} shows the cumulative mass distribution of the outburst ejecta.
We note that six outburst events, including 17P, 168P/Hergenrother, 332P, 217P/LINEAR, and two events at 15P, were brightened
down to 10 mag. Such events would be detectable with observation through inexpensive equipment such as $\sim$10-cm class telescopes.
Thus, we conjecture that outbursts $\lesssim$10 mag would
be detected almost completely if observed conditions allowed ground-based observers to make observations.

We fit the ejecta mass distribution with a power-law function for five objects (i.e., $\lesssim$10 mag events except 17P)
and obtained a power index of $\gamma$ = 0.45 $\pm$ 0.09. Although there might be no physical basis to support the power-law function
in Figure \ref{fig:cumM}, it is likely that the 17P outburst in 2007 was a very rare phenomenon to be detected within $\sim$8 years
because the occurrence is one order of magnitude
higher than that expected by the power function. In fact, the large discrepancy for 17P must be a bias of our analysis in which we counted the number of outbursts
since the epoch-making event. In addition, the power-law function fits well in the mass range of 6 $\times$ 10$^7$--1 $\times$ 10$^{9}$ kg
but deviates from the observed values in the mass range of $\lesssim$ 6 $\times$ 10$^7$ kg. Some outbursts might not be noticed
because of the faintness or weak contrast before and after outbursts. For these reasons, we consider the unbiased outbursts rate
in which the ejecta mass corresponds to 6 $\times$ 10$^7$--1 $\times$ 10$^{9}$, (hereafter referred to as "15P--class" outbursts).

We applied the Poisson distribution following \citet{Sonnett2011}:

\begin{eqnarray}
P(n)=\frac{(fCN)^n \exp(-fCN)}{n!},
\label{eq:poisson}
\end{eqnarray}

\noindent where $n$ and $f$ denote the number of outburst detections and incidence of outbursts,
$N$ is the number of observed samples, and $C$ ($\in$ [0, 1]) is the completeness of the survey.
The 90\% upper confidence limit on the incidence, $f_{90\%}$, is given by

\begin{eqnarray}
0.9=\frac{\int_0^{f_{90\%}} P(n)df}{\int_0^{1} P(n)df}~.
\label{eq:poisson}
\end{eqnarray}

Assuming that observers could detect 15P--class outbursts completely at the solar elongation of $\epsilon_\odot>$60\arcdeg,
we have $C$ = 0.67. Over eight years since October 2007, there were $n = 7$ 15P--class outbursts out of $N$ = 357 comets that passed
their perihelion. By solving the implicit Eq. (\ref{eq:poisson}), we obtained $f_{90\%}$ = 0.05, which suggests an expected number
$\langle n \rangle$ = $f_{90\%}CN$$\sim$12 in 8 years, 
or 1.5 times per year.
Therefore, we can express the unbiased cumulative and differential frequency of 15P--class outbursts $f_{ub}$ and $f'_{ub}$ as

\begin{eqnarray}
f_{ub}\left(>M\right) = \int^{\infty}_{M} f'_{ub}\left(m\right)~dm = A \left(\frac{M}{M_0}\right)^{-\gamma},
\label{eq:funbias}
\end{eqnarray}

\noindent where $A$ = 1.5 yr$^{-1}$ and $M_0$ = 6 $\times$ 10$^7$ kg are constants. 
We obtained an effective mass production rate of 15P--class outbursts of $\sim$ 3 $\times$ 10$^8$ kg year$^{-1}$ or $\sim$10 kg sec$^{-1}$, from (see appendix A)

\begin{eqnarray}
\langle m \rangle = \int^{M_2}_{M_1} m f'_{ub}(m)~dm~,
\end{eqnarray}

\noindent where $M_1$=6$\times$10$^7$ kg and $M_2$=1$\times$10$^9$ kg are lower and upper bounds on the power-law behavior of 15P--class outbursts, respectively.
We consider that the incidence would be underestimated because we optimistically assumed $C$ = 0.67. Some of the outbursts were missed owing to the crowded
region of stars or the full lunar phase. Even considering these factors, which may decrease $C$ by several factor, our result would obtain one order-of-magnitude accuracy
for the unbiased frequency. Therefore, it is reasonable to think  that 15P--class outbursts inject dust particles into the interplanetary space at a rate of $\gtrsim$10 kg sec$^{-1}$.

The mass is  $\sim$2 orders of magnitude less than the 
mass required to sustain the interplanetary dust cloud. This implies that 15P--class outburst events may not compensate the mass eroded by Poynting--Robertson
drag and other dynamical mechanisms onto the interplanetary dust. However, if we integrate $m f'_{ub}(m) dm$ up to the 17P-class ejecta mass (i.e., $M_2$=1$\times$10$^{12}$ kg),
the mass production rate from 15P--17P class events would be $\sim$500 kg sec$^{-1}$, although we are not sure whether the frequency follows a simple power-law function up to the 17P class.
These results suggest that large-scale cometary outbursts might contribute a significant fraction of the interplanetary dust source.

\section{SUMMARY}
\label{sec:summary}
We made an observation of 15P during the perihelion passage in 2015--2016, following a report of an outburst in the middle of 2015 December
and detected the second outburst on UT 2015 January 15.6--15.7, which was equivalent to the first outburst. The results of our analysis are summarized in the following points.

\begin{enumerate}
\item{ Gas consisting mostly of C$_{2}$ and CN expanded at a speed of 1,110 $\pm$ 180 km s$^{-1}$, which is slightly faster than the speeds
for other comets around 1 AU. The excess in speed can be explained by the large distance from the nucleus
($\approx 10^8$ km), where the gas flow velocity continues to increase.
}

\item{The dust ejecta accelerated up to a speed of 570 $\pm$ 40 km s$^{-1}$, which is comparable to the ejection speeds of 
17P and 332P ejecta. These consistent speeds would have resulted in the similar appearances of these outburst ejecta.
}

\item{We derived the total mass of dust ejecta as 10$^8$--10$^9$ kg ($a$ = 0.3 \micron--1 mm was assumed).
This mass is equivalent to that of the 2010 event at 332P but is three orders of magnitude smaller than the 2007 event at 17P.}

\item{The polarization degree was measured to 6.8 $\pm$ 0.2\% at the phase angle $\alpha$ = 43\arcdeg, which fell into the common
values of other comets. This similarity does not mean that outburst ejecta have similar polarimetric properties
because it was diffused out at the time of our polarimetric measurement.}

\item{Based on the immediate observation of dust ejecta for the second outburst, we derived a reliable estimate
of $\beta_\mathrm{max}$ = 1.6 $\pm$ 0.2. This value is consistent with the theoretical prediction for porous absorbing particles,
suggesting that such porous dust particles could remain inside the cometary nucleus and are released during the outburst.
}

\item{
The kinetic energy per unit mass (10$^4$ J kg$^{-1}$) is close to estimated values of
17P and 332P. In addition, the dust mass, speed, and kinetic energy are broadly comparable to the measured
values of the 2010 outburst at 332P. This may suggest that these three outbursts occurred by a similar mechanism.
}

\item{
From a survey of cometary
outbursts in publications in the SAO/NASA ADS, we estimated that 15P/Finlay--class outbursts
occur annually, injecting cometary materials into interplanetary space at a rate of $\gtrsim$10 kg sec$^{-1}$ or more.
}

\end{enumerate}

{\bf Acknowledgments}\\
This research, conducted at Seoul National University was supported by the Basic Science Research Program through the National Research
Foundation of Korea (NRF) funded by the Ministry of Education (NRF-2015R1D1A1A01060025, No. 2012R1A4A1028713). 
The acquisition of observation data was supported by the Optical and Near-infrared Astronomy Inter-University Cooperation Program.
We would like to thank Seiichi Yoshida for useful discussion and S. Goda for supporting observation at the Nayoro Observatory.
Y. G. Kwon is supported by Global Ph.D Fellowship Program through the NRF funded by the Ministry of Education (NRF-2015H1A2A1034260).

\appendix
\section{Statistics}
In this section, we prove the equations used in Section 4.2. Following \citet{Newman2005}, we consider a probability distribution of the form
\begin{eqnarray}
p(x) = \mathrm{Pr}(X = x) = Cx^{-\alpha}.
\end{eqnarray}
The cumulative distribution function of a power-law distributed variable is given by 
\begin{eqnarray}
P(x) = \mathrm{Pr}(X \geq x) = \int^{\infty}_{x} p\left(x'\right)~dx' = \left(\frac{x}{x_{\mathrm{min}}}\right)^{-\alpha+1},
\end{eqnarray}
where $x_{\mathrm{min}}$ is a lower bound on the power-law behavior. It's an identical form of Eq. (\ref{eq:funbias}), that is,
\begin{eqnarray}
f_{ub}\left(>M\right) = \int^{\infty}_{M} f'_{ub}\left(m\right)~dm = A \left(\frac{M}{M_0}\right)^{-\gamma},
\end{eqnarray}
where we consider $x$ and $x'$ correspond to $M$ and $m$, respectively.
Now we consider the mean value of our power-law distributed quantity $x$, given by
\begin{eqnarray}
\langle x \rangle = \int^{\infty}_{x_{\mathrm{min}}} x p\left(x\right)~dx = C\int^{\infty}_{x_{\mathrm{min}}} x^{-\alpha+1}~dx = \frac{C}{2-\alpha}\left[x^{-\alpha+2}\right]^{\infty}_{x_{\mathrm{min}}}.
\end{eqnarray}
It also corresponds that we integrate $m f'_{ub}(m) dm$ to obtain the mean mass production rate of 15P class outbursts at a given mass range. We can write the equation:
\begin{eqnarray}
\langle m \rangle = \int^{M_1}_{M_0} m f'_{ub}(m)~dm,
\end{eqnarray}
where $M_0$ and $M_1$ are lower and upper bounds on the power-law behavior, respectively.


\clearpage
\begin{deluxetable}{lrrrrrrrrl}
\tablecaption{Observation and event summary\label{tab:journal}}
\tablewidth{450pt}
\tabletypesize{\scriptsize}
\tablehead{\colhead{Median UT} & \colhead{Telescope} & \colhead{Filter} & \colhead{$N^a$} & \colhead{$T_\mathrm{tot}^b$} & \colhead{r$_\mathrm{h}^c$} & \colhead{$\Delta^d$} & \colhead{$\alpha^e$} & \colhead{$f_\mathrm{T}^f$} & \colhead{Note}}
\startdata
\\
(2014-12-16.0) & \ldots & \ldots & \ldots & \ldots & 0.989 &\ldots & \ldots & 346.0 & 1st outburst$^\dag$\\
\\
2014-12-23.398 & OAO 0.5-m & {\sl g}$'$, $R_\mathrm{C}$, $I_\mathrm{C}$ & 58 & 58.0 & 0.977 & 1.443 & 42.8 & 355.1& Close to the Mars\\
2014-12-23.402 & NHAO 2-m & r$'$ & 20 & 20.0 & 0.977 & 1.443 & 42.8 & 355.1& Close to the Mars\\
2014-12-23.433 & IAO 1.05-m & {\sl g}$'$, $R_\mathrm{C}$, $I_\mathrm{C}$ & 18 & 54.0 & 0.977 & 1.442 & 42.8 & 355.1& Close to the Mars\\
2014-12-25.400 & OAO 0.5-m & {\sl g}$'$, $R_\mathrm{C}$, $I_\mathrm{C}$ & 62 & 62.0 & 0.976 & 1.434 & 43.1 & 357.8\\
2014-12-25.406 & HHO 1.5-m & $R_\mathrm{C}$ & 17 & 17.0 & 0.976 & 1.434 & 43.2 & 357.8 & Polarimetry\\
2014-12-26.375 & NO 1.6-m & $R_\mathrm{C}$ & 17 & 17.0 & 0.976 & 1.431 & 43.3 & 359.1 & Polarimetry\\
2014-12-26.419 & OAO 0.5-m & {\sl g}$'$, $R_\mathrm{C}$, $I_\mathrm{C}$ & 17 & 17.0 & 0.976 & 1.430 & 43.3 & 359.1\\
\\
(2014-12-27)  & \ldots & \ldots & \ldots & \ldots & 0.976 &\ldots & \ldots & 0.0 & Perihelion\\
\\
2014-12-27.369 & NHAO 2-m & $R_\mathrm{C}$ & 15 & 12.5 & 0.976 & 1.427 & 43.5 & 0.4\\
2014-12-27.405 & OAO 0.5-m & {\sl g}$'$, $R_\mathrm{C}$, $I_\mathrm{C}$ & 72 & 72.0 & 0.976 & 1.427 & 43.5 & 0.5\\
2014-12-29.402 & OAO 0.5-m & {\sl g}$'$, $R_\mathrm{C}$, $I_\mathrm{C}$ & 57 & 57.0 & 0.977 & 1.420 & 43.8 & 3.1\\
2014-12-30.404 & OAO 0.5-m & {\sl g}$'$, $R_\mathrm{C}$, $I_\mathrm{C}$ & 68 & 68.0 & 0.977 & 1.417 & 43.9 & 4.5\\
2015-01-03.409 & OAO 0.5-m & {\sl g}$'$, $R_\mathrm{C}$, $I_\mathrm{C}$ & 72 & 72.0 & 0.982 & 1.406 & 44.4 & 9.8\\
2015-01-06.400 & OAO 0.5-m & {\sl g}$'$, $R_\mathrm{C}$, $I_\mathrm{C}$ & 47 & 47.0 & 0.988 & 1.399 & 44.6 & 13.7\\
2015-01-07.409 & OAO 0.5-m & {\sl g}$'$, $R_\mathrm{C}$, $I_\mathrm{C}$ & 63 & 63.0 & 0.990 & 1.398 & 44.7 & 15.1 & Weak eruption\\
2015-01-08.418 & OAO 0.5-m & {\sl g}$'$, $R_\mathrm{C}$, $I_\mathrm{C}$ & 20 & 20.0 & 0.993 & 1.396 & 44.8 & 16.4\\
2015-01-10.437 & IAO 1.05-m & {\sl g}$'$, $R_\mathrm{C}$, $I_\mathrm{C}$ & 15 & 15.0 & 0.999 & 1.394 & 44.9 & 19.0\\
2015-01-11.430 & NHAO 2-m & $R_\mathrm{C}$ & 3 & 3.0 & 1.002 & 1.393 & 44.9 & 20.2\\
2015-01-11.436 & IAO 1.05-m & {\sl g}$'$, $R_\mathrm{C}$, $I_\mathrm{C}$ & 27 & 27.0 & 1.002 & 1.393 & 44.9 & 20.3\\
2015-01-13.420 & OAO 0.5-m & {\sl g}$'$, $R_\mathrm{C}$, $I_\mathrm{C}$ & 84 & 84.0 & 1.009 & 1.392 & 44.9 & 22.8\\
\\
(2015-01-15.6--15.7)  & \ldots & \ldots & \ldots & \ldots & 1.017 &\ldots & \ldots & 25.5 & 2nd outburst\\
\\
2015-01-16.448 & IAO 1.05-m & {\sl g}$'$, $R_\mathrm{C}$, $I_\mathrm{C}$ & 6 & 3.0 & 1.021 & 1.393 & 44.9 & 26.5 &\\
2015-01-17.434 & OAO 0.5-m & {\sl g}$'$, $R_\mathrm{C}$, $I_\mathrm{C}$ & 54 & 54.0 & 1.025 & 1.394 & 44.9 & 27.7\\
2015-01-18.400 & OAO 0.5-m & {\sl g}$'$, $R_\mathrm{C}$, $I_\mathrm{C}$ & 3 & 3.0 & 1.030 & 1.394 & 44.8 & 28.9\\
2015-01-19.411 & OAO 0.5-m & {\sl g}$'$, $R_\mathrm{C}$, $I_\mathrm{C}$ & 25 & 25.0 & 1.034 & 1.396 & 44.8 & 30.1\\
2015-01-23.458 & IAO 1.05 m & {\sl g}$'$, $R_\mathrm{C}$, $I_\mathrm{C}$ & 16 & 48.0 & 1.055 & 1.403 & 44.5 & 34.8\\
2015-01-24.424 & OAO 0.5-m & {\sl g}$'$, $R_\mathrm{C}$, $I_\mathrm{C}$ & 65 & 65.0 & 1.060 & 1.406 & 44.4 & 35.9 & Close to Moon\\
2015-01-24.440 & IAO 1.05 m & {\sl g}$'$, $R_\mathrm{C}$, $I_\mathrm{C}$ & 14 & 14.0 & 1.061 & 1.406 & 44.4 & 36.0 & Close to Moon\\
2015-01-25.436 & IAO 1.05 m & {\sl g}$'$, $R_\mathrm{C}$, $I_\mathrm{C}$ & 4 & 4.0 & 1.066 & 1.408 & 44.3 & 37.1\\
2015-01-30.427 & NHAO 2-m & $R_\mathrm{C}$ & 30 & 25.0 & 1.097 & 1.426 & 43.6 & 42.5\\
2015-01-31.429 & OAO 0.5-m & {\sl g}$'$, $R_\mathrm{C}$, $I_\mathrm{C}$ & 29 & 29.0 & 1.103 & 1.430 & 43.4 & 43.6\\
2015-02-02.422 & OAO 0.5-m & {\sl g}$'$, $R_\mathrm{C}$, $I_\mathrm{C}$ & 47 & 47.0 & 1.117 & 1.440 & 43.1 & 45.7\\
2015-02-08.451 & OAO 0.5-m & {\sl g}$'$, $R_\mathrm{C}$, $I_\mathrm{C}$ & 113 & 113.0 & 1.160 & 1.475 & 41.9 & 51.6\\
2015-02-13.440 & OAO 0.5-m & {\sl g}$'$, $R_\mathrm{C}$, $I_\mathrm{C}$ & 45 & 90.0 & 1.198 & 1.511 & 40.7 & 56.2\\
2015-02-18.445 & OAO 0.5-m & {\sl g}$'$, $R_\mathrm{C}$, $I_\mathrm{C}$ & 43 & 86.0 & 1.239 & 1.553 & 39.5 & 60.5\\
2015-02-20.465 & IAO 1.05 m & {\sl g}$'$, $R_\mathrm{C}$, $I_\mathrm{C}$ & 13 & 39.0 & 1.256 & 1.571 & 39.0 & 62.2\\
2015-03-16.375 & OAO 1.88 m & $R_\mathrm{C}$ & 10 & 28.0 & 1.472 & 1.852 & 32.3 & 78.8\\
\\
\enddata
\tablenotetext{a}{Number of exposures}
\tablenotetext{b}{Total exposure time [min]}
\tablenotetext{c}{Heliocentric distance [AU]}
\tablenotetext{d}{Geocentric distance [AU]}
\tablenotetext{e}{Solar phase angle [degrees]}
\tablenotetext{f}{True anomaly [degrees]}
\tablenotetext{\ddag}{Because  accurate time was not known for the first outburst, we quoted $r_\mathrm{h}$ and $f_\mathrm{T}$ at UT 00:00 on the possible day.}
\end{deluxetable}
\clearpage


\begin{table}
  \caption{Dust model parameters}
  \begin{center}
    \begin{tabular}{lllll}
\hline
Parameter   & Input Values & Best-fit (Envelope) & Best-fit (Tail + Coma)
     & Unit\\
\hline
$u_1$ & 0.1--0.9 with 0.1 interval & 0.4 $\pm$ 0.3 & 0.55 $\pm$ 0.1 & --\\
$q$ & 3.0--4.5 with 0.1 interval & 3.8 $\pm$ 0.1 & 3.8 $\pm$ 0.1 & --\\
$\beta_\mathrm{max}$ & 1.0--2.5  with 0.1 interval &
	     1.6 & 1.6  & --\\
$\beta_\mathrm{min}$ & 0.5, 0.3, 0.1, 0.01, 0.001
	 & 0.3 & 6$\times$10$^{-4}$ (fixed) & --\\
$V_0$ & 150--600 with 30 interval & 450 $\pm$ 30 & 330$^{+60}_{-30}$ & m s$^{-1}$\\
$\sigma_v$ & 0--0.5 with 0.1 interval & 0.2 $\pm$ 0.1 & 0.4 $\pm$ 0.1 & --\\
$\omega$ & 5--60 with 5 interval & 35 $\pm$ 10 & 30 $\pm$ 10 & degree\\
\hline
    \end{tabular}
  \end{center}
 \label{tab:parameter}
\end{table}
\clearpage


\begin{table}
\footnotesize
  \caption{Comparison of 15P, 322P, and 17P Outbursts}
  \begin{center}
     \begin{tabular}{lrrrr}
\hline
Quantity     & 15P & 332P & 17P & References \tablenotemark{17}\\
\hline
$a$\tablenotemark{1} & 3.488& 3.090     & 3.621 & (1)\\
$e$\tablenotemark{2} & 0.720 & 0.489     & 0.432 & (1)\\
$i$\tablenotemark{3} & 6.799 & 9.378     & 19.090 & (1)\\
$q_p$\tablenotemark{4} & 0.976 & 1.579     & 2.057  & (1)\\
$T_J$        & 2.620 & 3.010     & 2.859   & (1)\\
$r_h$\tablenotemark{5} & 0.99, 1.02 & 1.59      & 2.44 & (2), (3)\\
$R_N$\tablenotemark{6} & 0.9 & $<$1.9   & 2.1 & (4), (2), (5)\\
$\Delta t_p$\tablenotemark{7} & -11, +19 &+20       & +172  & (3)\\
$m_R(1,1,0)$\tablenotemark{8} & 7.2, 5.6  & 6.0 & -1.1 & (2)\\
$C_\mathrm{c}$\tablenotemark{9}  & $>$2.6$\times$10$^{10}$, 1.4$\times$10$^{11}$& 1.0$\times$10$^{11}$ &
	      7.1$\times$10$^{13}$ & (2), (6)\\
$t_{rise}\tablenotemark{10}$ & $<$3 & $\approx$1 & 1.2$\pm$0.3 & (7)\\
$t_{fade}\tablenotemark{11}$&  --           & 70 & 50 & (8)\\
$M_d$\tablenotemark{12}   & $>$2$\times$10$^{8}$, 7$\times$10$^{8}$ & 5$\times$10$^{8}$ &
	      (4--8)$\times$10$^{11}$ & \\
$V_{max}$\tablenotemark{13}   & 570$\pm$40 & 500$\pm$40 & 554$\pm$5 &(2), (10)\\
$E_k$\tablenotemark{14}     & 7.0$\times$10$^{12}$ for 2nd & 5.0$\times$10$^{12}$ &
	      (1--6)$\times$10$^{15}$ & (2), (10)\\
$E_k/M_d$\tablenotemark{15} & 1$\times$10$^{4}$ & 1$\times$10$^{4}$  & $10^4$--$10^5$
	      & (2), (7), (10)\\
\hline
     \end{tabular}
   \tablenotetext{1}{Semi-major axis in AU.}
   \tablenotetext{2}{Eccentricity.}
   \tablenotetext{3}{Inclination in degree.}
   \tablenotetext{4}{Perihelion distance in AU.}
   \tablenotetext{5}{Heliocentric distance at the time of outburst in AU.}
   \tablenotetext{6}{Radius of nucleus in km.}
   \tablenotetext{7}{Onset time after perihelion passage in days.}
   \tablenotetext{8}{Absolute $R_\mathrm{C}$--band magnitude.}
   \tablenotetext{9}{Total cross-section of dust cloud in m$^2$.}
   \tablenotetext{10}{Rise time in days.}
   \tablenotetext{11}{Fade time when the magnitude decreased by 4 mag in days.}
   \tablenotetext{12}{Ejecta mass in kg.}
   \tablenotetext{13}{Maximum speed of ejecta in m s$^{-1}$.}
   \tablenotetext{14}{Kinetic energy in J.}
   \tablenotetext{15}{Kinetic energy per unit mass in J kg$^{-1}$.}
   \tablenotetext{16}{We obtained these by using images taken at Kiso Observatory.}
   \tablenotetext{17}{References: (1) ssd.jpl.nasa.gov, (2) \citet{Ishiguro2014}, (3) \citet{Hsieh2010}, (4) \citet{Fernandez2013}, (5) \citet{Stevenson2014}, (6) \citet{Ishiguro2013}, (7)  \citet{Li2011}, (8) \citet{Stevenson2012}, (9) \citet{Lin2009}, (10) \citet{Ishiguro2016}}
  \end{center}
 \label{tab:comparison}
\end{table}
\clearpage

\begin{landscape}
\begin{table}[htbp]
\footnotesize
  \caption{List of outbursts at JFCs and ETCs since 2007 October}
\begin{center}
\begin{tabular}{llccccccccc}
\hline
Name & Date & $r_\mathrm{h}$ & $\Delta$ & $\alpha$ & $m_\mathrm{-}$\tablenotemark{1} & $m_\mathrm{+}$\tablenotemark{2} & $m_\mathrm{-}(1,1,0)$\tablenotemark{3} & $m_\mathrm{+}(1,1,0)$\tablenotemark{4} & $C_\mathrm{c}$\tablenotemark{5} & Reference\tablenotemark{6}\\
\hline
205P/Giacobini & 2015 Sep 26 & 1.92 & 1.57 & 32 & $\sim$20 & $\sim$14 & 16.5 & 10.5 & 1.6$\times 10^9$ & (1)\\
17P/Holmes & 2015 Jan 26 & 2.99 & 2.38 & 26 & 19.2 & 16.8 & 14.0 & 11.6 & 5.1$\times 10^8$ & (2)\\
15P/Finlay   & 2015 Jan 15 & 1.02 & 1.39	& 45 & 11.1 & 8.0 & 11.1 & 5.6 & 1.4 $\times 10^{11}$ & (3)\\
                    & 2015 Jan 7    & 0.99 & 1.40 & 45 & 13.2 & 12.7 & 10.9 & 10.4 & 6.8$\times 10^8$ & (3)\\
                    & 2014 Dec      & 0.99 & 1.48 & 41 & -- & 9.4 & -- & 7.2 & 2.6$\times 10^{10}$ & (3)\\
63P/Wild 1 & 2013 May 16 & 1.98 & 1.58 & 30 & 15.6 & 13.5 & 12.0 & 10.0 & 2.2$\times 10^9$ & (4)\\
168P/Hergenrother & 2012 Aug--Dec & 1.42 & 0.43 & 13 & 15.2 & 8.0 & 15.8 & 8.6 & 9.1$\times 10^9$ & (5)\\
P/2010 V1 & 2010 Nov 2 & 1.59 & 2.34 & 19 & -- & 9.5 & -- & 6.0 & 1.0$\times 10^{11}$ & (6)\\
103P/Hartley 2 & 2010 Aug--Nov & 1.06--1.56 & 0.12--0.70 & 29--59 & -- & -- & -- & -- & -- &(7)\\
81P/Wild 2 & 2010 Apr--Aug & 1.70--2.21 & 0.70--1.85 & 4.5--27.1 &	-- & -- & -- & -- & 2.2$\times 10^9$ & (8)\\
P/2010 H2 & 2010 Apr & 3.11 & 2.13 & 5 & $>$20 & 12.6 & 15.7 & 8.3 & 1.2$\times 10^{10}$ & (9)\\
217P/LINEAR & 2009 Oct	 & 1.31 & 0.61 & 47 & 11.3 & 9.3 & 10.1 & 8.1 & 1.2$\times 10^{10}$ & (10)\\
199P/Shoemaker 4 & 2008 Aug 3 & 3.41 & 3.35 & 17 & 17.8 & 14.5 & 14.1 & 8.6 & 8.9$\times 10^9$ & (11)\\
6P/d'Arrest & 2008 Aug & - & - & - & - & - & - & - & - &									(11)\\
17P/Holmes & 2008 Jan 5 & 4.18 & 3.28 & 6 & 19.8 & 19.3 & 13.9 & 13.4 & 4.5$\times 10^7$ & (12)\\
                    & 2007 Nov 12 & 2.51 & 1.62 & 13 & 12.8 & 12.6 & 9.2 & 9.0 & 1.1$\times 10^9$ & (13)\\
                    & 2007 Oct 23 & 2.44 & 1.64 & 17 & 17 & 2.5 & 13.4 & -1.1 & 7.1$\times 10^{13}$ & (14)\\
\hline
\end{tabular}
\tablenotetext{1}{Pre-outburst apparent magnitudes}
\tablenotetext{2}{Post-outburst apparent magnitudes}
\tablenotetext{3}{Pre-outburst absolute magnitudes calculated with Eq. (\ref{eq:abs-mag})}
\tablenotetext{4}{Post-outburst absolute magnitudes calculated with Eq. (\ref{eq:abs-mag})}
\tablenotetext{5}{Cross-section calculated with Eq. (\ref{eq:CrossSection}) assuming the geometric albedo $p_\mathrm{R}$=0.04.}
\tablenotetext{6}{References: (1) S. Yoshida, private communication, (2) \citet{Kwon2016}, (3) This work, (4) \citet{Opitom2013}, (5) \citet{Sekanina2014}, (6) \citet{Ishiguro2013}, (7)\citet{Milani2013}, (8) \citet{Bertini2012}, (9) \citet{Vales2010}, (10) \citet{Sarugaku2010}, (11) Miles 2009, (12) \citet{Miles2010}, (13) \citet{Stevenson2012}, (14) \citet{Li2011}}
\end{center}
\end{table}
\end{landscape}
\clearpage

\clearpage

\clearpage

\clearpage

\clearpage

\clearpage

\clearpage

\clearpage

\clearpage

\end{document}